\documentclass[12pt]{article}
\usepackage{amstex,amscd}

\newtheorem{xdefinition}{Definition}
\newtheorem{xtheorem}[xdefinition]{Theorem}
\newtheorem{xlemma}[xdefinition]{Lemma}
\newtheorem{xcorollary}[xdefinition]{Corollary}
\newtheorem{xexample}[xdefinition]{Example}
\newenvironment{definition}{\begin{xdefinition}\rm}%
{\hspace*{\fill}\raisebox{-1pt}{\boldmath$\Box$}\end{xdefinition}}
\newenvironment{theorem}{\begin{xtheorem}\rm}{\end{xtheorem}}
\newenvironment{lemma}{\begin{xlemma}\rm}{\end{xlemma}}
\newenvironment{corollary}{\begin{xcorollary}\rm}{\end{xcorollary}}
\newenvironment{example}{\begin{xexample}\rm}%
{\hspace*{\fill}\raisebox{-1pt}{\boldmath$\Box$}\end{xexample}}
\newenvironment{proof}{\begin{trivlist}\item[]{\flushleft\bf Proof }}
     {\hspace*{\fill}\raisebox{-1pt}{\boldmath$\Box$}\end{trivlist}}
\newcommand{\definitionref}[1]{Definition~\ref{def:#1}}
\newcommand{\theoremref}[1]{Theorem~\ref{thm:#1}}
\newcommand{\lemmaref}[1]{Lemma~\ref{lm:#1}}

\newcommand{\exampleref}[1]{Example~\ref{ex:#1}}

\newcommand{\sectlabel}[1]{\label{sect.#1}}
\newcommand{\sect}[1]{\mbox{Section~\ref{sect.#1}}}
\newcommand{\eqlabel}[1]{\label{eq.#1}}
\newcommand{\eq}[1]{Equation~(\ref{eq.#1})}

\makeatletter                    
\def\title#1{\gdef\@title{#1\vskip.5em}}
\def\affil#1{\\{\protect\small\sl #1\par}}
\gdef\@abstract{}
\long\def\abstract#1{\begin{center}{\bf Abstract}
    \begin{small}\begin{quote}{#1}\end{quote}\end{small}\end{center}}
\makeatother

\title{Efficient Quantum Transforms}
\author{Peter H{\o}yer%
\thanks{Supported in part by the {\sc esprit} Long Term Research
Programme of the EU under project number 20244 ({\sc alcom-it}).
Current address: D\'ept.~IRO, Universit\'e de Montr\'eal.
Email: hoyer$\mathchar"40$IRO.UMontreal.CA.}
\affil{Odense University%
\thanks{Department of Mathematics and Computer Science,
Odense University, Campusvej~55, \mbox{DK--5230} Odense~M, Denmark.
Email: u2pi$\mathchar"40$imada.ou.dk.}}}
\date{February~11, 1997}

\def\mb#1{\mbox{$#1$}}           
\def\Co{\Bbb C}                  
\def\integer{\Bbb Z}             
\def\ket#1{\mbox{$|#1\rangle$}}  
\def\span#1{\mbox{$\langle #1\rangle$}}
\DeclareMathSymbol{\leqslant}   {\mathrel}{AMSa}{"36}  
\def\subgroup{\leqslant}         
\def\th#1{\mbox{$#1$--th}}       
\newcommand{\sis}[3]{\sideset{^{(#1)\!}}{_{#2}^{#3}}}
\newcommand{\pst}[2]{\vphantom{#1}\smash[t]{#1^{#2}}}

\begin{document}
\maketitle
\thispagestyle{empty}

\vspace{-.5cm} \abstract{Quantum mechanics requires the operation of
quantum computers to be unitary, and thus makes it important to have
general techniques for developing fast quantum algorithms for computing
unitary transforms.  A~quantum routine for computing a generalized
Kronecker product is given.  Applications include re-development of the
networks for computing the Walsh-Hadamard and the quantum Fourier
transform.  New networks for two wavelet transforms are given.  Quantum
computation of Fourier transforms for non-Abelian groups is defined.
A~slightly relaxed definition is shown to simplify the analysis and the
networks that computes the transforms.  Efficient networks for computing
such transforms for a class of metacyclic groups are introduced.
A~novel network for computing a Fourier transform for a group used in
quantum error-correction is also given.}

\section{Introduction}
The quantum computational version of the discrete Fourier transform is
without doubt the most important transform developed for quantum
computing so far.  It~is in the heart of all quantum computational
issues discussed until now.  All main quantum algorithms, including
Shor's celebrated factoring algorithm~\cite{Shor97} and Grover's
searching algorithm~\cite{Grover96}, use~it as a subroutine.  All known
relativized separation results for quantum computation are based on
quantum algorithms that use the discrete Fourier
transform~\cite{BB94,BV97,Simon94,BBBV97}.  Fundamental concepts of
quantum error-correction rely on it; see for
example~\cite{CRSS97,CS96,Gottesman96,Steane96a,Steane96b}.  It~is,
in~conclusion, the most important single routine for obtaining an
insight in the previous work done in quantum computing.  Still, this
seemingly simple transform is not yet fully understood.

When referring to the discrete Fourier transform, one often does not
refer to a single transform, but rather to a family of transforms.  For
any positive integer~$n$ and any $n$--dimensional complex vector
space~$V_n$, one defines a discrete Fourier transform~$F_n$ (see for
example~\cite{VanLoan92}).  More generally, given~$r$ positive
integers~$\{n_i\}_{i=1}^r$, and $r$~complex vector
spaces~$\{V_i\}_{i=1}^r$, where $V_i$ is of dimension~$n_i$, one defines
a discrete Fourier transform, denoted $F_{n_1} \otimes \dots \otimes
F_{n_r}$, for the tensor product space $V_1 \otimes \dots \otimes V_r$.
(See \sect{qft} for details.)

Earlier, the quantum versions of the discrete Fourier transforms were
defined, but {\em efficient\/} quantum networks where only known for few
of them.  Currently, efficient networks implementing~$F_n$ exactly have
been found for all smooth
integers~$n$~\cite{Cleve94,Coppersmith94,DJ92,Grigoriev96,Shor97},
where~$n$ is considered smooth if all its prime factors are less than
$\log^c(n)$ for some constant~$c$~\cite{Shor97}.  Furthermore, any
discrete Fourier transform can be efficiently approximated to any degree
of accuracy by some quantum circuit~\cite{BL95,Kitaev95}.

Common for the networks discussed above are that their description has
been taken from the point of view that the transforms were to be
implemented by quantum networks.  In~this paper, we discuss quantum
computation not as opposed to classical (perhaps parallelized)
computation, but more as a variant.  We~believe that the fundamental
object is the unitary transform, which then can be considered a quantum
or a classical (reversible) algorithm.  With this point of view, the
problem of finding an efficient algorithm implementing a given unitary
transform~$U$, reduces to the problem of factorizing~$U$ into a small
number of ``sparse'' unitary transforms such that those sparse
transforms should be known to be efficiently implementable.  As~an
example of this, we show that the quantum networks implementing the
quantum versions of the discrete Fourier transforms can be very easily
derived from the mathematical descriptions of their classical
counterparts.

More generally, we consider a new tool for finding quantum networks
implementing any given unitary transform~$U$.  We~show that if~$U$ can
be expressed as a certain generalized Kronecker product (defined below)
then, given efficient quantum networks implementing each factor in this
expression, we also have an efficient quantum network implementing~$U$.
The expressive power of the generalized Kronecker product includes
several new transforms.  Among these are the two wavelet transforms: the
Haar transform~\cite{Haar10} and Daubechies'~$D^4$
transform~\cite{Daubechies88}, and we are thus able to devise new
quantum networks that compute these transforms.

There exists a group theoretical interpretation of the discrete Fourier
transforms which establishes a bijective correspondence between this
family of transforms and the set of finite Abelian groups.  To~further
demonstrate the power of the generalized Kronecker products, we use them
to give a simple re-development of the quantum Fourier transforms for
Abelian groups.  More interestingly, the Fourier transforms can be
generalized to arbitrary finite non-Abelian groups (see for
example~\cite{MR95} for an introduction), and we give a definition of
what it means that a quantum network computes such transforms.
Moreover, we give a slightly relaxed definition where we only compute a
Fourier transform up to phase factors.  Classically, the idea of
relating a Fourier transform for a group to one of its subgroups has
proven to be very useful~\cite{MR95}, and we show that this carries over
to quantum computers.  We~apply these ideas to give new networks for
quantum computing Fourier transforms for the quaternionic group and for
a class of metacyclic groups.

Since Shor demonstrated that quantum error-correction is possible by
given an explicit nine-bit code~\cite{Shor95}, several new classes of
quantum codes have been developed.  See for example
\cite{CRSS97,CS96,Gottesman96,Steane96a,Steane96b} for some of the many
results.  Many of these are stabilizer codes, which are subgroups of a
certain non-Abelian group~$E_n$ (defined in \sect{errorcorr}).  For that
group, we also give a simple and efficient network for computing a
Fourier transform, again using the framework of generalized Kronecker
products.

Independently of this work, Robert Beals has found quantum networks
implementing Fourier transforms for the symmetric groups~\cite{Beals97}.
A~challenging open question related to that result is whether it can be
used to find a polynomial quantum circuit solving the famous graph
isomorphism problem.

\section{Generalized Kronecker products}\sectlabel{gkp}
All matrices throughout this paper are finite.  Matrices are denoted by
bold capital letters and tuples of matrices by calligraphic letters.
Indices of tuples and row and column indices of matrices and vectors are
numbered starting from zero; the \th{(i,j)} element of~$\mathbf A$ is
referred to as~$a_{ij}$.  A~single integer as subscript on a unitary
matrix denotes its dimension, e.g., we let \mb{\mathbf I_q}~denote the
\mb{(q \times q)} identity matrix.  The transpose of~$\mathbf A$ is
denoted by~\mb{\mathbf A^t}.  Recall that a square matrix is unitary if
it is invertible and its inverse is the complex conjugate of its
transpose.  The complex conjugate of a number~$c$ is denoted by
$\overline c$.

\begin{definition}\label{def:kro}
Let $\mathbf A$ be a \mb{(p \times q)} matrix and $\mathbf C$ a \mb{(k
\times l)} matrix.  The {\em left\/} and {\em right Kronecker product\/}
of $\mathbf A$ and $\mathbf C$ are the \mb{(pk \times ql)} matrices
\begin{equation*}
\begin{bmatrix} \mathbf A c_{00}  &
\mathbf A c_{01}  & \dots & \mathbf A c_{0,l-1} \\ \mathbf A c_{10}  &
 \mathbf A c_{11}
 & \dots & \mathbf A c_{1,l-1}  \\ \vdots & \vdots & & \vdots\\
\mathbf A c_{k-1,0} & \mathbf A c_{k-1,1}  & \dots & \mathbf A c_{k-1,l-1} 
\\ \end{bmatrix} \;\; \text{ and } \;\;
\begin{bmatrix} a_{00} \mathbf C &
a_{01} \mathbf C & \dots & a_{0,q-1} \mathbf C\\ a_{10} \mathbf C & a_{11}
\mathbf C & \dots & a_{1,q-1} \mathbf C \\ \vdots & \vdots & & \vdots\\
a_{p-1,0} \mathbf C & a_{p-1,1} \mathbf C & \dots & a_{p-1,q-1} \mathbf C
\\ \end{bmatrix},
\end{equation*}
respectively.
\end{definition}
We~denote the left Kronecker product by $\mathbf A \otimes_L \mathbf C$
and the right Kronecker product by $\mathbf A \otimes_R \mathbf C$.
When some property holds for both definitions, we use \mb{\mathbf A
\otimes \mathbf C}.  Note that the Kronecker product is a binary matrix
operator as opposed to the tensor product which is binary operator
defined for algebraic stuctures like modules.  The Kronecker product can
be generalized in different ways; see for example \cite{FA77}
and~\cite{RM89}.  In~this paper, we use (an even further generalized
version of) the generalized Kronecker product discussed in~\cite{FA77},
defined as follows.
 
\begin{definition}\label{def:gkp}
Given two tuples of matrices, a $k$--tuple \mb{{\mathcal A} = (\mathbf
A^i)_{i=0}^{k-1}} of \mb{(p \times q)} matrices and a $q$--tuple
${\mathcal C} = (\mathbf C^i)_{i=0}^{q-1}$ of \mb{(k \times l)}
matrices, the {\em generalized right Kronecker product\/} is the~\mb{(pk
\times ql)} matrix \mb{\mathbf D = \mathcal A \otimes_R \mathcal C}
where
\[ d_{ij} = d_{uk+v,xl+y} = a_{ux}^v c_{vy}^x \] 
with $0 \leq u < p$, $0 \leq v < k$, $0 \leq x < q$, and $0 \leq y < l$.
\end{definition}

The generalized right Kronecker product can be found from the standard
right Kronecker product by, for each sub-matrix \mb{a_{ux} \mathbf C} in
\definitionref{kro} substituting it with the following sub-matrix
\begin{equation*}\begin{bmatrix}
a_{ux}^0 c_{00}^x & a_{ux}^0 c_{01}^x & \dots & a_{ux}^0 c_{0,l-1}^x \\
a_{ux}^1 c_{10}^x & a_{ux}^1 c_{11}^x & \dots & a_{ux}^1 c_{1,l-1}^x \\
 \vdots           & \vdots           &       & \vdots\\
a_{ux}^{k-1} c_{k-1,0}^x 
     & a_{ux}^{k-1} c_{k-1,1}^x & \dots & a_{ux}^{k-1} c_{k-1,l-1}^x \\
\end{bmatrix}.
\end{equation*}
 
The generalized {\em left\/} Kronecker product is the \mb{(pk \times
ql)} matrix \mb{\mathbf D = \mathcal A \otimes_L \mathcal C}
where the \th{(i,j)} entry holds the value
\[d_{ij} = d_{up+v,xq+y} = a_{vy}^u c_{ux}^y\]
with $0 \leq u < k$, $0 \leq v < p$, $0 \leq x < l$, and $0 \leq y < q$.

As~for standard Kronecker products, we let \mb{\mathcal A \otimes
\mathcal C} denote either of the two definitions.  If~the matrices
\mb{{\mathbf A}^i = \mathbf A} are all identical, and also \mb{\mathbf
C^i = \mathbf C}, the generalized Kronecker product \mb{\mathcal A
\otimes \mathcal C} reduces to the standard Kronecker product
\mb{\mathbf A \otimes \mathbf C}.  Denote by \mb{\mathcal A \otimes
\mathbf C} the generalized Kronecker product of a $k$--tuple
\mb{\mathcal A} of \mb{(p \times q)} matrices, and a $q$--tuple
\mb{\mathcal C} of identical \mb{(k \times l)} matrices~$\mathbf C$.
Denote \mb{\mathbf A \otimes \mathcal C} similarly.
\begin{example}
\label{ex:haar:equations}
\begin{multline*}\left( 
\begin{bmatrix}1&\phantom{-}1\\1&-1\end{bmatrix},
\begin{bmatrix}1&0\\0&1\end{bmatrix} \right) \otimes_R
\begin{bmatrix}1&\phantom{-}1\\1&-1\end{bmatrix}\\
= \begin{bmatrix}\begin{tabular}{cc|cc} $1 \cdot 1$&$1 \cdot 1 $&$ 1
\cdot 1$&$ 1 \cdot 1$\\ $1 \cdot 1$&$1 \cdot (-1) $&$ 0 \cdot 1$&$ 0
\cdot (-1)$\\\hline $1 \cdot 1$&$1 \cdot 1 $&$ (-1) \cdot 1$&$ (-1)
\cdot 1$\\ $0 \cdot 1$&$0 \cdot (-1) $&$ 1 \cdot 1$&$ 1 \cdot (-1)$
\end{tabular} \end{bmatrix}
= \begin{bmatrix}1&\phantom{-}1&\phantom{-}1&\phantom{-}1\\
1&-1&\phantom{-}0&\phantom{-}0\\1&\phantom{-}1&-1&-1\\
0&\phantom{-}0&\phantom{-}1&-1\end{bmatrix}
\end{multline*}
 
\begin{multline*}\left( 
\begin{bmatrix}1&\phantom{-}1\\1&-1\end{bmatrix},
\begin{bmatrix}1&0\\0&1\end{bmatrix} \right) \otimes_L
\begin{bmatrix}1&\phantom{-}1\\1&-1\end{bmatrix}\\
= \begin{bmatrix}\begin{tabular}{cc|cc}
$1 \cdot 1$&$1  \cdot 1$&$1 \cdot 1$&$1 \cdot 1$\\
$1 \cdot 1$&$(-1) \cdot 1$&$1 \cdot 1$&$(-1) \cdot 1$\\\hline
$1 \cdot 1$&$0 \cdot 1$ &$1 \cdot (-1)$&$0 \cdot (-1)$\\
$0 \cdot 1$&$1 \cdot 1$&$0 \cdot (-1)$&$1 \cdot (-1)$
\end{tabular} \end{bmatrix}
= \begin{bmatrix}1&\phantom{-}1&\phantom{-}1&\phantom{-}1\\
1&-1&\phantom{-}1&-1\\1&\phantom{-}0&-1&\phantom{-}0\\
0&\phantom{-}1&\phantom{-}0&-1\end{bmatrix}.
\end{multline*}
\end{example}

To~analyze generalized Kronecker products we need the {\em shuffle
permutation matrix} of dimension \mb{(mn \times mn)}, denoted $\Pi_{mn}$
as shorthand for $\Pi_{(m,n)}$, defined~by
\[ \pi_{rs} = \pi_{dn+e,d'm+e'} = \delta_{de'} \delta_{d'e}, \]
where \mb{0 \leq d,e' < m; \; 0 \leq d',e <n}, and~$\delta_{xy}$
denotes the Kronecker delta function which is zero if~\mb{x \neq y}, and
one otherwise.  It~is unitary and satisfies $\Pi_{mn}^{-1} = \Pi_{mn}^t
= \Pi_{nm}$.

Given two tuples of matrices, $k$--tuple $\mathcal A = ({\mathbf
A}^i)_{i=0}^{k-1}$ of \mb{(p \times r)} matrices and $k$--tuple
$\mathcal C = (\mathbf C^i)_{i=0}^{k-1}$ of \mb{(r \times q)} matrices,
let \mb{\mathcal A \mathcal C} denote the $k$--tuple where the \th{i}
entry is the \mb{(p \times q)} matrix \mb{\mathbf A^i \mathbf C^i}, $0
\leq i < k$.  For any $k$--tuple $\mathcal A$ of matrices, let
\mb{{\text{Diag}}(\mathcal A)} denote the direct sum
\mb{\bigoplus_{i=0}^{k-1} \mathbf A^i} of the matrices $\mathbf A^0,
\dots, \mathbf A^{k-1}$.  The generalized Kronecker products satisfy the
following important Diagonalization Theorem of~\cite{FA77}.

\begin{theorem}\label{thm:diag}[Diagonalization Theorem]
Let~\mb{\mathcal A = ({\mathbf A}^i)_{i=0}^{k-1}} be a $k$--tuple
of~\mb{(p \times q)} matrices and \mb{\mathcal C = (\mathbf
C^i)_{i=0}^{q-1}} a $q$--tuple of \mb{(k \times l)} matrices.  Then
\begin{align}\eqlabel{gkp:fac:r}
\!\mathcal A \otimes_R \mathcal C &= 
  \big(\Pi_{pk} \, {\text{Diag}}(\mathcal A) \, \Pi_{kq}\big) \times
  {\text{Diag}}(\mathcal C)\\ \eqlabel{gkp:fac:l}
\!\mathcal A \otimes_L \mathcal C &= 
  {\text{Diag}}(\mathcal A) \times
  \big(\Pi_{kq} \, {\text{Diag}}(\mathcal C) \, \Pi_{ql}\big).
\end{align}
\end{theorem}

\begin{corollary}
Let~\mb{\mathcal A = ({\mathbf A}^i)_{i=0}^{k-1}} be a $k$--tuple
of~\mb{(p \times q)} matrices and \mb{\mathcal C = (\mathbf
C^i)_{i=0}^{q-1}} a $q$--tuple of \mb{(k \times l)} matrices.  Then
\begin{eqnarray*}
\mathcal A \otimes_R \mathcal C &=& \Pi_{pk} \,
    \big(\mathcal A \otimes_L \mathcal C\big) \, \Pi_{lq}\\
\mathcal A \otimes_L \mathcal C &=& \Pi_{kp} \,
    \big(\mathcal A \otimes_R \mathcal C\big) \, \Pi_{ql}.
\end{eqnarray*}
\end{corollary}

Until now, we have not assumed anything about the dimension of the
involved matrices.  In~the next theorem, we assume that the matrices
involved are square matrices.  The theorem is easily proven from the
Diagonalization Theorem.  For any $k$--tuple~$\mathcal A = ({\mathbf
A}^i)_{i=0}^{k-1}$ of invertible matrices, let $\mathcal A^{-1}$ denote
the $k$--tuple where the \th{i} entry equals the inverse of~$\mathbf
A^i$, $0 \leq i < k$.

\begin{corollary}
Let $\mathcal A, \mathcal C$ be $m$--tuples of \mb{(n \times n)}
matrices, and $\mathcal D, \mathcal E$ be $n$--tuples of \mb{(m \times
m)} matrices.  Then
\begin{equation}\eqlabel{sep}
\mathcal A \mathcal C   \otimes \mathcal D \mathcal E
   =  \big(\mathcal A   \otimes \mathbf  I_m\big) \times 
      \big(\mathcal C   \otimes \mathcal D\big) \times 
      \big(\mathbf  I_n \otimes \mathcal E\big).
\end{equation}
Furthermore, if the matrices in the tuples $\mathcal A \text{ and }
\mathcal C$ are invertible, then
\begin{gather*}
\big(\mathcal A \otimes_R \mathcal C\big)^{-1} 
  = \Pi_{nm} \, \big(\mathcal C^{-1} \otimes_R 
      \mathcal A^{-1}\big) \, \Pi_{mn}
  = \mathcal C^{-1} \otimes_L \mathcal A^{-1}\\
\big(\mathcal A \otimes_L \mathcal C\big)^{-1} 
  = \Pi_{mn} \, \big(\mathcal C^{-1} \otimes_L 
      \mathcal A^{-1}\big) \, \Pi_{nm}
  = \mathcal C^{-1} \otimes_R \mathcal A^{-1}\\
\text{If $\mathcal A$ and $\mathcal C$ are unitary, then 
  so is $\mathcal A \otimes \mathcal C$.}
\end{gather*}
\end{corollary}

\section{Quantum routines}\sectlabel{routines}
In~this section, we give a method for constructing a quantum network for
computing any given generalized Kronecker product.  A~primary
application of this method is as a tool to find a quantum network of a
given unitary matrix.  As~two examples, we use it to develop quantum
networks for computing two wavelet transforms, the Haar transform and
Daubechies' $\mathbf D^4$~transform.

As~our quantum computing model, we~adopt the now widely used quantum
gates arrays~\cite{BBC+95,Berthiaume97}.  Let \mb{\tau : \ket{u,v}
\mapsto \ket{u, v \oplus u}} denote the two-bit exclusive-or operation,
and $\cal U$ the set of all one-bit unitary operations.
Following~\cite{BBC+95}, by a {\em basic operation\/} we mean either a
$\cal U$~operation or the $\tau$~operation.  The collection of basic
operations is universal for quantum networks in the sense that any
finite quantum network can be approximated with arbitrary precision by a
quantum network~$Q$ consisting only of gates implementing such
operations~\cite{Deutsch89,Yao93,BBC+95}.  

Define the one-bit unitary operations
\[\mathbf X = \begin{bmatrix}0&1\\1&0\end{bmatrix} \quad
\mathbf Z = \begin{bmatrix}1&\phantom{-}0\\0&-1\end{bmatrix} \quad
\mathbf Y = \begin{bmatrix}0&-1\\1&\phantom{-}0\end{bmatrix} \quad
\mathbf W = \frac1{\sqrt 2}\begin{bmatrix}1&\phantom{-}1\\1&-1
\end{bmatrix}.\] 
Given a unitary matrix~$\mathbf C$, let $\Lambda((j,x),(k,\mathbf C))$
denote the transform where we apply~$\mathbf C$ on the \th{k} register
if and only if the \th{j} register equals~$x$.  Given an $n$--tuple
\mb{\mathcal C = ({\mathbf C}^i)_{i=0}^{n-1}} of unitary matrices, let
\mb{\Lambda((j,i),(k,\mathbf C^i))_i} denote $\Lambda((j,n-1),(k,\mathbf
C^{n-1})) \cdots \Lambda((j,0),(k,\mathbf C^0))$.  Given a \th{k} root
of unity, say $\omega$, let $\Phi(\omega)$ denote the unitary transform
given by $\ket{u} \ket{v} \mapsto \omega^{uv} \ket{u}\ket{v}$.  If~the
first register holds a value from~$\Bbb Z_n$, and the second holds a
value from~$\Bbb Z_m$, then $\Phi(\omega) = \Phi_{(n,m)}(\omega)$ can be
implemented in $\Theta(\lceil \log n \rceil \lceil \log m \rceil)$ basic
operations~\cite{Cleve94,Coppersmith94,Kitaev95}.

{\bf Quantum shuffle transform.}  For every \mb{m > 1}, let the
operation $\text{DM}_m$ perform the unitary transform \mb{\ket{k}\ket{0}
\mapsto \ket{k \text{ div } m} \ket{k \text{ mod } m}}.  Let \text{SWAP}
denote the unitary transform \mb{\ket{u} \ket{v} \mapsto \ket{v}
\ket{u}}.  Then $\Pi_{mn}$ can be implemented on a quantum computer by
one application of $\text{DM}_m$, one swap operation, and one
application of $\text{DM}_n^{-1}$,
\[\Pi_{mn} \equiv \text{DM}_n^{-1} \:\; \text{SWAP} \:\;
  \text{DM}_m.\]

{\bf Quantum direct sum.}  Let $\mathcal C$ be an $n$--tuple of $(m
\times m)$ unitary matrices.  It~is not difficult to see that
${\text{Diag}}(\mathcal C)$ can be implemented as follows,
\[{\text{Diag}}(\mathcal C) \equiv \text{DM}_m^{-1} \:\;
  \Lambda((1,i),(2,\mathbf C^i))_i \:\; \text{DM}_m.\] In~general, the
time to compute the direct sum is proportional to the sum of the
computation times of each of the conditional $\mathbf C^i$~transforms.
However, if parts of these can be applied in quantum parallel, this
improves the running time.

{\bf Quantum Kronecker product.}  Let~$\mathcal A$ be an $m$--tuple of
$(n \times n)$ unitary matrices and $\mathcal C$ an $n$--tuple of $(m
\times m)$ unitary matrices.  By~the Diagonalization Theorem, the
generalized Kronecker product can be applied by applying two direct sums
and two shuffle transforms.  Removing cancelling terms we get
\begin{align}
\mathcal A \otimes_R \mathcal C &\equiv \text{DM}_m^{-1} \:
 \Lambda((2,i),(1,\mathbf A^i))_i \; \Lambda((1,i),(2,\mathbf C^i))_i
 \:\; \text{DM}_m \eqlabel{rkp-impl} \\ \mathcal A \otimes_L
 \mathcal C &\equiv \text{DM}_n^{-1} \: \Lambda((1,i),(2,\mathbf
 A^i))_i \; \Lambda((2,i),(1,\mathbf C^i))_i \:\; \text{DM}_n.
\end{align}
Thus, an application of a generalized right Kronecker product can be
divided up into the following four steps: in the first step, we apply
$\text{DM}_m$.  In~the second step, we apply the controlled $\mathbf
C^i$ transforms on the second register, and in the third step, the
controlled $\mathbf A^i$ transforms on the first register.  Finally, in
the last step, we apply $\text{DM}_m^{-1}$ to the result.

\begin{example}
Let $\mathcal A$ be a 4--tuple of $(2 \times 2)$ unitary matrices,
and $\mathcal C$ a 2--tuple of $(4 \times 4)$ unitary matrices. The
generalized Kronecker product $\mathcal A \otimes_R \mathcal C$
can be implemented by a quantum network as follows.
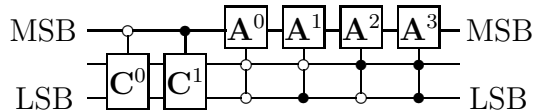
\begin{center}\leavevmode\hbox{
\setlength{\unitlength}{0.09mm}
\begin{picture}(720,160)(0,80)
\put(70,100){\line(1,0){30}}
\put(160,100){\line(1,0){25}}
\put(245,100){\line(1,0){52.5}}
\put(312.5,100){\line(1,0){70}}
\put(397.5,100){\line(1,0){70}}
\put(482.5,100){\line(1,0){70}}
\put(567.5,100){\line(1,0){52.5}}
\put(70,150){\line(1,0){30}}
\put(160,150){\line(1,0){25}}
\put(245,150){\line(1,0){52.5}}
\put(312.5,150){\line(1,0){70}}
\put(397.5,150){\line(1,0){70}}
\put(482.5,150){\line(1,0){70}}
\put(567.5,150){\line(1,0){52.5}}
\put(70,200){\line(1,0){52.5}}
\put(275,200){\line(-1,0){137.5}}
\put(335,200){\line(1,0){25}}
\put(420,200){\line(1,0){25}}
\put(505,200){\line(1,0){25}}
\put(590,200){\line(1,0){30}}
\put(130,165){\line(0,1){27.5}}
\put(215,165){\line(0,1){27.5}}
\put(305,157.5){\line(0,1){17.5}}
\put(305,107.5){\line(0,1){35}}
\put(390,157.5){\line(0,1){17.5}}
\put(390,107.5){\line(0,1){35}}
\put(475,157.5){\line(0,1){17.5}}
\put(475,107.5){\line(0,1){35}}
\put(560,157.5){\line(0,1){17.5}}
\put(560,107.5){\line(0,1){35}}
\put(130,125){\makebox(0,0){\framebox(60,80){$\mathbf C^0$}}}
\put(215,125){\makebox(0,0){\framebox(60,80){$\mathbf C^1$}}}
\put(305,205){\makebox(0,0){\framebox(60,60){$\mathbf A^0$}}}
\put(390,205){\makebox(0,0){\framebox(60,60){$\mathbf A^1$}}}
\put(475,205){\makebox(0,0){\framebox(60,60){$\mathbf A^2$}}}
\put(560,205){\makebox(0,0){\framebox(60,60){$\mathbf A^3$}}}
\put(305,100){\circle{15}}\put(305,150){\circle{15}}
\put(390,100){\circle*{15}}\put(390,150){\circle{15}}
\put(475,100){\circle{15}}\put(475,150){\circle*{15}}
\put(560,100){\circle*{15}}\put(560,150){\circle*{15}}
\put(130,200){\circle{15}}
\put(215,200){\circle*{15}}
\put(7,100){\makebox(0,0){LSB}}
\put(5,200){\makebox(0,0){MSB}}
\put(680,200){\makebox(0,0){MSB}}
\put(678,100){\makebox(0,0){LSB}}
\end{picture}}
\end{center}

The dots and the circles represent control-bits: if the values with the
dots are one, and if the values with the circles are zero, the transform
is applied, otherwise it is the identity map. The least (most)
significant bit is denoted by LSB (MSB).  Note that most of the
transforms are orthogonal, and thus the gates commute.  Note also that,
following the ideas of Griffiths and Niu in~\cite{GN95}, a semi-classical
generalized Kronecker product transform can be defined.
\end{example}

\subsection{Quantum wavelet transforms}\sectlabel{qwt}
A~main application of the Diagonalization Theorem is as a tool to find
quantum networks computing large unitary matrices.  Suppose we have
factorized a unitary matrix~$\mathbf U$ via a generalized Kronecker
product into a product of some simpler matrices.  Then, if we have
quantum networks for computing these simpler transforms, we also have a
quantum network for computing~$\mathbf U$ by applying the methods
developed above.  We~give two examples of this technique; in both cases
implementing a wavelet transform.

\begin{example}\label{ex:haar}
The Haar wavelet transform~\cite{Haar10}, $\mathbf H$, can be defined
using the generalized Kronecker product as follows.
\begin{equation}\eqlabel{haar}
\begin{split}
\mathbf H_2 &= \mathbf W\\
\mathbf H_{2^{n+1}} &= \Pi_{2,2^n} \times \big((\mathbf H_{2^n}, 
    \mathbf I_{2^n}) \otimes_R \mathbf W\big), \quad n=1,2,\dotsc
\end{split}
\end{equation}
Applying the decomposition of the generalized right Kronecker product
given in \eq{rkp-impl}, we immediately obtain an efficient quantum
circuit for computing the Haar transform.  Let~$\mathbf S_{2^{n+1}}$
denote the bit-shift transform given by $\ket{b_n \dots b_0} \mapsto
\ket{b_0 b_n \dots b_1}$.  This transform efficiently
implements~$\Pi_{2,2^n}$.  For~$n=3$, the quantum circuit defined by
\eq{haar} is given below.  Here the two $\mathbf S$~transforms are the
bit-shift transforms of the appropriate dimensions.

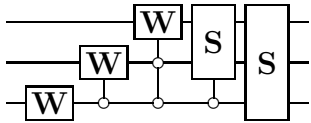
\begin{center}\leavevmode\hbox{
\setlength{\unitlength}{0.09mm}
\begin{picture}(580,135)(0,125)
\put(70,100){\line(1,0){30}}
\put(70,160){\line(1,0){110}}
\put(70,220){\line(1,0){190}}
\put(134,100){\makebox(0,0){\framebox(68,48){$\mathbf W$}}}
\put(214,160){\makebox(0,0){\framebox(68,48){$\mathbf W$}}}
\put(294,220){\makebox(0,0){\framebox(68,48){$\mathbf W$}}}
\put(214,100){\circle{15}}
\put(294,100){\circle{15}}
\put(376,100){\circle{15}}
\put(294,160){\circle{15}}
\put(376,190){\makebox(0,0){\framebox(62,108){$\mathbf S$}}}
\put(455,160){\makebox(0,0){\framebox(62,168){$\mathbf S$}}}
\put(168,100){\line(1,0){38.5}}
\put(248,160){\line(1,0){38.5}}
\put(221.5,100){\line(1,0){65}}
\put(301.5,100){\line(1,0){67}}
\put(301.5,160){\line(1,0){42.5}}
\put(383.5,100){\line(1,0){40.5}}
\put(424,160){\line(-1,0){17}}
\put(424,220){\line(-1,0){17}}
\put(345,220){\line(-1,0){17}}
\put(486,100){\line(1,0){30}}
\put(486,160){\line(1,0){30}}
\put(486,220){\line(1,0){30}}
\put(214,107.5){\line(0,1){28.5}}
\put(294,167.5){\line(0,1){28.5}}
\put(376,107.5){\line(0,1){28.5}}
\put(294,107.5){\line(0,1){45}}
\end{picture}}
\end{center}
\end{example}

By \eq{sep}, we can rewrite the recursive definition in \eq{haar} as
\begin{equation*}
\mathbf H_{2^{n+1}} = \Pi_{2,2^n} \times \big((\mathbf H_{2^n},
    \mathbf I_{2^n}) \otimes_R \mathbf I_2\big) \times 
    \big(\mathbf I_{2^n} \otimes_R \mathbf W\big).
\end{equation*}
We~refer to the right-most factor in this factorization as the {\em
scaling matrix\/} of dimension $(2^{n+1} \times 2^{n+1})$ for the Haar
wavelet transform.  In~general, given any family of unitary
matrices~$\{\mathbf D_{2^i}\}_{i \geq i_0}$, define a family of unitary
transforms $\{\mathbf U_{2^i}\}_{i \geq i_0}$ as follows,
\begin{equation*}
\begin{split}
\mathbf U_{2^{i_0}} &= \mathbf D_{2^{i_0}}\\
\mathbf U_{2^{n+i_0}} &= \Pi_{2,2^{n+i_0-1}} \times 
    \big((\mathbf U_{2^{n+i_0-1}}, \mathbf I_{2^{n+i_0-1}}) 
    \otimes_R \mathbf I_2\big) \times \mathbf D_{2^{n+i_0}}, 
    \quad n=1,2,\dotsc
\end{split}
\end{equation*}
We~refer to the $\mathbf D_i$~matrices as a family of {\em scaling
matrices}, and the family of $\mathbf U_i$~matrices as a {\em wavelet
transform}.  Suppose that we have a family of efficient quantum networks
for computing a given family of scaling matrices.  Then, as in
\exampleref{haar}, we also have efficient quantum networks for computing
the associated wavelet transform.  The next example gives a
factorization of the scaling matrices used in Daubechies'~$D^4$ wavelet
transform~\cite{Daubechies88}.

\begin{example}
Let $m \geq 4$ be an even integer, and let
\begin{equation*}
k_{0/1} = \frac{3 \pm \sqrt 3}{4 \sqrt 2} \quad \text{ and } \quad
k_{2/3} = \frac{1 \mp \sqrt 3}{4 \sqrt 2}.
\end{equation*}
Daubechies' $\mathbf D^4_m$ scaling matrix~\cite{Daubechies88} of
dimension $(m \times m)$ is the matrix with
\begin{equation*}
d_{ij} = \begin{cases}k_{j-i+x} &\text{ $i$ is even}\\
             (-1)^j k_{2+i-j-x} &\text{ $i$ is odd,}\end{cases}
\end{equation*}
where 
\begin{equation*}
x = \begin{cases} 4 &\text{ if $i \geq m-2$ and $j<2$}\\
		  0 &\text{ otherwise},\end{cases}
\end{equation*}
and $k_l=0$ if $l<0 \text{ or if } l>3$. 

Let~$\mathbf P_m$ be the $(m \times m)$ permutation matrix which
subtracts two if the input is odd, \mbox{i.e.}, $p_{ij}=1$ if $i=j$ and
$i$ is even, or if $i+2 \equiv j \text{ (mod~$m$)}$ and $i$ is odd.  Let
$\mathbf C_0$ and~$\mathbf C_1$ denote the two one-bit unitary
operations,
\begin{equation*}
\mathbf C_0 = 2 \begin{pmatrix}k_3&-k_2\\k_2&\phantom{-}k_3\end{pmatrix}
\quad \text{ and } \quad \mathbf C_1 
    = \frac12 \begin{pmatrix}k_0/k_3&1\\1&k_1/k_2\end{pmatrix}.
\end{equation*}
The scaling matrix $\mathbf D^4_m$ can then be factorized using two
Kronecker products
\begin{equation}\eqlabel{D4fact}
\mathbf D^4_m = \big(\mathbf I_{m/2} \otimes_R \mathbf C_1\big) 
  \times \mathbf P_m \times \big(\mathbf I_{m/2} \otimes_R
  \mathbf C_0\big).
\end{equation}
Set~$n = \lceil \log m \rceil$.  The permutation transform~$\mathbf P_m$
can be implemented in $\Theta(n)$ basic operations~\cite{VBE96}.  Each
of the other two factors on the right hand side of \eq{D4fact} can be
implemented in one basic operation.  Thus, $\mathbf D^4_m$ can be
implemented in $\Theta(n)$ basic operations.  We~remark that we have not
been able to find this factorization of~$\mathbf D_m^4$ elsewhere in the
literature---despite the fact that using it compared to straightforward
use of~$\mathbf D^4_m$ saves $m$~additions in the classical case.
Furthermore, note that $\left(\mathbf I_{m/2} \otimes_R \mathbf
C_1\right) \times \left(\mathbf I_{m/2} \otimes_R \mathbf C_0\right) =
\mathbf I_{m/2} \otimes_R \mathbf W$, which is the scaling matrix used
in the Haar transform.
\end{example}

\section{Group representations and quantum Fourier\\ transforms}
\sectlabel{qft}
In~the rest of this paper, $G$~will denote a finite group, written
multiplicative with identity~$e$, and~$\eta$ the order of~$G$.  Let~$\Co
G$ denote the complex group algebra of~$G$.  Let~${\cal
B}_{\text{time}}$ denote the standard basis of~$\Co G$, that is,
$\{g_1,\dots,g_\eta\}$, and let $(u,v) = \sum_{g \in G} u(g)
\overline{v(g)}$ denote the natural inner product in~$\Co G$.  Let
\mb{\text{GL}_d(\Co)} denote the multiplicative group of \mb{(d \times
d)} invertible matrices with complex entries.  We~start by reviewing
some basic facts from the theory of linear representations of finite
groups.  For a general introduction to group representation theory, see
for example \cite{CR62} or~\cite{Serre77}.

A~complex matrix {\em representation}~$\rho$ of~$G$ is a
group-homomorphism \mb{\rho : G \rightarrow \text{GL}_d(\Co)}.  The
dimension \mb{d=d_\rho} is called the {\em degree\/} or {\em
dimension\/} of the representation~$\rho$.  Two representations,
$\rho_1$~and~$\rho_2$, of degree~$d$ are {\em equivalent\/} if there
exists an invertible matrix~$A \in \text{GL}_d(\Co)$ such that
\mb{\rho_2(g) = A^{-1} \rho_1(g) A} for all \mb{g \in G}.
A~representation \mb{\rho : G \rightarrow \text{GL}_d(\Co)} is {\em
irreducible\/} if there is no non-trivial subspace of~$\Co^d$ which is
invariant under \mb{\rho(g)} for all \mb{g \in G}, and it is {\em
unitary\/} if \mb{\rho(g)} is unitary for all \mb{g \in G}.  For every
representation there exists an equivalent unitary representation.  Up~to
equivalence, there are only a finite number of irreducible
representations, say~$\nu$, of~$G$.  This number equals the number of
distinct conjugate classes of~$G$.

Let~\mb{{\cal R} = \{\rho^1,\dots,\rho^\nu\}} be a complete set of
inequivalent, irreducible and unitary representations of~$G$ with $d_i$
equal to the degree of~$\rho^i$.  For any representation~$\rho \in {\cal
R}$, the vector \mb{\rho_{kl} \in \Co G} defined by considering the
\th{(k,l)} entry of~\mb{\rho(g)} for each \mb{g \in G} is called a {\em
matrix coefficient\/} of~$\cal R$.  The inner product of two matrix
coefficients of~$\cal R$ is non-zero if and only if they are equal.  For
each matrix coefficient~$\rho_{kl}$, let $b_{\rho,k,l}$ denote the
normalized matrix coefficient, and let ${\cal B}_{\text{freq}} =
\{b_{\rho,k,l}\}$ denote the set of orthonormalized matrix coefficients.
Since one can show that the degrees~$d_i$ of the representations $\rho_i
\in {\cal R}$ satisfy the relation \mb{\sum_{i=1}^\nu d_i^2 = \eta}, it
follows that ${\cal B}_{\text{freq}}$ is an orthonormal basis of the
vector space~$\Co G$.

The linear operator~$F_G$ on~$\Co G$ which maps a vector $v \in \Co G$
given in the standard basis~${\cal B}_{\text{time}}$ to its
representation $\Hat v \in \Co G$ in basis~${\cal B}_{\text{freq}}$ is
called the {\em Fourier transform\/} for~$\Co G$ on~$\cal R$.  Each
entry of~$\Hat v$, denoted~$\Hat v(\rho_{kl})$ or just $\Hat \rho_{kl}$,
is called a {\em Fourier coefficient\/} of~$v$ (on~$\cal R$).

In~the recent years, many new exciting results have been found
concerning the computation of Fourier transforms for finite groups on
classical computers---see~\cite{MR95} for a nice survey.  In~this paper,
we consider the computation of Fourier transforms on quantum computers.
Since quantum mechanics requires the operation of the computer to be
unitary, our definition of a Fourier transform given above is slightly
more strict than the most common used definitions for the classical
case.

We~now define what it means that a quantum circuit computes a Fourier
transform.  Let~$F_G$ be a Fourier transform for~$\Co G$ on~$\cal R$.
Let~$E_{\text{time}}: {\cal B}_{\text{time}} \rightarrow \integer_\eta$
and~$E_{\text{freq}}: {\cal B}_{\text{freq}} \rightarrow \integer_\eta$
be two bijections.  These functions induce an ordering on ${\cal
B}_{\text{time}}$ and~${\cal B}_{\text{freq}}$, respectively.
Let~$E:{\cal B}_{\text{time}} \cup {\cal B}_{\text{freq}} \rightarrow
\integer_\eta$ denote the extension of~$E_{\text{time}}$
and~$E_{\text{freq}}$.  We~say that~$E$ is an {\em encoding\/} for the
linear transform~$F_G$.  With respect to~$E$, $F_G$~can be viewed as a
matrix~$\mathbf F_G$ in~$\text{GL}_\eta (\Co)$.  This matrix is unitary
by construction, and thus there exists a quantum circuit computing
it~\cite{Deutsch89}.  We~say that the circuit {\em computes\/} $F_G$
with respect to~$E$.

Given a $k$--tuple of complex numbers of unit norm, $(\phi_i)_{i=1}^k$,
let $\phi = \text{diag}(\phi_i) \in \text{GL}_k (\Co)$ denote the
unitary diagonal matrix with $\phi_i$ at the \th{i} diagonal entry.
Let~$F_G$ be a Fourier transform for~$\Co G$ on~$\cal R$, $E$~an
encoding for~$F_G$, and $\mathbf F_G$ the resulting unitary matrix.
We~say that a quantum circuit {\em computes $F_G$ up to phase factors\/}
(with respect to~$E$) if there exists a unitary diagonal matrix~$\phi
\in \text{GL}_\eta (\Co)$ such that the circuit computes \mb{\mathbf
F_G^\phi = \phi \, \mathbf F_G}.  Given a network that computes $F_G$ up
to phase factors, we can obtain a quantum circuit for computing~$F_G$
exactly by applying first $\mathbf F_G^\phi$ and then the unitary
transform \mb{\phi^{-1} = \phi^\star}.

Note the dependencies of the set~$\cal R$ and the encoding~$E$ of ${\cal
B}_{\text{time}}$ and~${\cal B}_{\text{freq}}$ in the above definitions.
A~Fourier transform for~$\Co G$ is defined only with respect to~$\cal
R$.  A~quantum circuit computing the Fourier transform is, in addition,
defined with respect to an encoding~$E$ of the basis-elements in ${\cal
B}_{\text{time}}$ and~${\cal B}_{\text{freq}}$.

The quantum computation time of~$F_G$ with respect to~$\cal R$ and the
encoding~$E$ is defined as the minimum number of basic operations in any
quantum circuit computing~$\mathbf F_G$, and it is denoted
by~\mb{\text{QT}(G)({\cal R},E)}.  The quantum computation time of a
Fourier transform for~$\Co G$, denoted~$\text{QT}(G)$, is defined as the
minimum of~$\text{QT}(G)({\cal R},E)$ over all possible choices of $\cal
R$ and~$E$.

In~the rest of this paper, $\cal R$ denotes a complete set of
inequivalent, irreducible and unitary representations of~$G$.

\section{Quantum Fourier transforms for cyclic groups}
\sectlabel{qft-cyclic} As~an introductionary example, consider the
problem of quantum computing the discrete Fourier transform.  We~start
by developing an efficient quantum routine for computing the discrete
Fourier transform using the generalized Kronecker product discussed
in~\sect{gkp}.  Then, we review a group theoretical interpretation of
the transform which relates it to the cyclic groups.
 
The discrete Fourier transform for a quantum computer is defined as
follows. For any positive integer~$n$, let
\begin{equation}\eqlabel{dft}
\mathbf F_n \ket{x} 
  = \frac1{\sqrt n} \sum_{y=0}^{n-1} \omega_n^{xy} \ket{y},
\end{equation}
for each $x=0,\dots,n-1,$ where \mb{\omega_n = \exp(2\pi \sqrt{-1} /n)}
is the principal \th{n} root of unity.  The unitary Fourier
transform~$\mathbf F_{nm}$ can be defined from $\mathbf F_n$
and~$\mathbf F_m$ using a generalized Kronecker product
\begin{equation}\eqlabel{qfourier}
\mathbf F_{nm} = \Pi_{nm} \times \big(\mathbf F_n \otimes_L \mathbf
    I_m\big) \times \big((\mathbf D_{nm}^s)_{s=0}^{m-1} \otimes_L
    \mathbf I_m\big) \times \big(\mathbf I_n \otimes_L \mathbf F_m\big)
\end{equation}
where $\mathbf D_{nm}^s = \text{diag}(\omega^{si})$ for $0 \leq s < m$
and $\omega = \omega_{nm}$.  \eq{qfourier} is referred to as a {\em
radix--$n$ splitting\/} in~\cite{VanLoan92}, where a proof of the
identity can be found.
 
\eq{qfourier} gives an efficient quantum routine for computing $\mathbf
F_{nm}$ from $\mathbf F_n, \mathbf F_m$, and $\mathbf D_{nm}^s$.
Interestingly, the resulting routine obtained this way is the same as
the one found by Cleve~\cite{Cleve94} using a direct method.  The
transform $((\mathbf D_{nm}^s)_s \otimes_L \mathbf I_m)$ is a special
application of the $\Phi$~transform defined in \sect{routines} and is
thus easily applied.  For powers of~2, the computation of~$\mathbf
F_{2^n}$ uses $\Theta(n^2)$ basic operations~\cite{Coppersmith94}.

We~now review the well-known group theoretical correspondence to the
discrete Fourier transform.  Let~$G = \Bbb Z_n$ be the cyclic group of
order~$n$.  For Abelian groups all irreducible representations are
one-dimensional, and hence equivalent representations are equal.  There
are $n$~distinct representations, ${\cal R} =
\{\zeta^0,\dots,\zeta^{n-1}\}$, given~by
\[\zeta^i(j) = [\overline \omega_n^{ij}] \, 
  \text{ for every $j \in \Bbb Z_n$}.\] 

The collection of normalized matrix coefficients are ${\cal
B}_{\text{freq}} = \{b_{\zeta^0,1,1},\dots,b_{\zeta^{n-1},1,1}\}$ where
$(b_{\zeta^i,1,1},j) = \frac{1}{\sqrt n} \overline \omega_n^{ij}$ for
all $b_{\zeta^i,1,1} \in {\cal B}_{\text{freq}}$ and $j \in {\cal
B}_{\text{time}}$.  Hence, for all $b_{\zeta^i,1,1} \in {\cal
B}_{\text{freq}}$ we have $b_{\zeta^i,1,1} = \frac1{\sqrt n} \sum_{j \in
{\cal B}_{\text{time}}} \overline \omega_n^{ij} \, j$.  Thus, by
choosing the encoding~$E$ given by $E_{\text{time}}(j)=j$ and
$E_{\text{freq}}(b_{\zeta^i,1,1})=i$, respectively, the quantum circuit
defined by~\eq{dft} is seen to compute the Fourier transform for the
cyclic group~$\Bbb Z_n$ with respect to~$E$.  We~remark that it is
possible also to give a group theoretical interpretation of the
decomposition given by~\eq{qfourier}; see for example~\cite{MR95} for
details.

\subsection{Direct product groups}\sectlabel{product}
Suppose we are given quantum networks (as black-boxes) for quantum
computing Fourier transforms for the group algebras $\Co G_1$ and~$\Co
G_2$.  Consider the problem of computing a Fourier transform for the
direct product group algebra \mb{\Co G = \Co(G_1 \times G_2)}.
Classically, this problem has a very simple solution.  In~this section,
we show that this carries over in the quantum circuit model.

Let $G_1$ and~$G_2$ be finite groups of order $\eta_1$ and~$\eta_2$,
respectively.  Our first task is to establish a specific
isomorphism~$\varphi$ between the algebras $\Co G_1 \times \Co G_2$ and
$\Co(G_1 \times G_2)$.  Let ${\cal B}_{\text{time}}^i$ denote the
standard basis of~$\Co G_i$, \mb{i=1,2}, and let ${\cal B}_{\text{time}}
= \{(g_1,g_2) : g_i \in G_i, \: i=1,2\}$ denote the standard basis of
\mb{\Co (G_1 \times G_2)}.  Let $\Co G_1 \otimes \Co G_2$ denote the
tensor product algebra of $\Co G_1$ and~$\Co G_2$, and $\varphi : \Co
G_1 \otimes \Co G_2 \rightarrow \Co (G_1 \times G_2)$ the natural
algebra isomorphism defined by
\[\varphi (g_1 \otimes g_2) = (g_1,g_2) \quad \left(g_1 \otimes g_2 \in 
   {\cal B}_{\text{time}}^1 \otimes {\cal B}_{\text{time}}^2\right).\]
With these definitions, we can write
\begin{equation}
{\cal B}_{\text{time}} = \varphi\left({\cal B}_{\text{time}}^1
  \otimes {\cal B}_{\text{time}}^2\right).
\end{equation}

Let ${\cal R}_i$ be a complete set of inequivalent, irreducible and
unitary representations of $G_i$, \mb{i=1,2}.  We~need the following
lemma from representation theory.
\begin{lemma}\label{lm:product}
Let $G_i$ and ${\cal R}_i$ be given as above, $i=1,2$.  Then
\begin{equation*}
{\cal R} = {\cal R}_1 \otimes_R {\cal R}_2
     = \{ \rho_1 \otimes_R \rho_2 : \rho_i \in {\cal R}_i, \: i=1,2\}
\end{equation*}
is a complete set of inequivalent, irreducible and unitary
representations of~$G=G_1 \times G_2$.
\end{lemma}

Let ${\cal B}_{\text{freq}}^i$ denote the set of orthonormalized matrix
coefficients of~${\cal R}_i$, \mb{i=1,2}, and ${\cal B}_{\text{freq}}$
the set of orthonormalized matrix coefficients of~$\cal R$, where $\cal
R$ is given as in \lemmaref{product}.  By~that lemma, it follows by
straightforward algebra that
\begin{equation}
{\cal B}_{\text{freq}} = \varphi\left({\cal B}_{\text{freq}}^1
  \otimes {\cal B}_{\text{freq}}^2\right).
\end{equation}

Having established the above isomorphism, we now state the main result
from representation theory to be used in this section.  Essentially, it
reduces the problem of computing a Fourier transform for $\Co(G_1 \times
G_2)$ to those of computing Fourier transforms for $\Co G_1$ and~$\Co
G_2$.

\begin{theorem}\label{thm:product1}
Let~$F_1$ and~$F_2$ be Fourier transforms for $\Co G_1$ and~$\Co G_2$ on
${\cal R}_1$ and ${\cal R}_2$, respectively.  Define the linear
transform~$F'_G$ over $\Co G_1 \otimes \Co G_2$~by
\[F'_G(g_1 \otimes g_2) = F_1(g_1) \otimes F_2(g_2).\]
Then $F_G = \varphi F'_G \varphi^{-1}$ is a Fourier transform for
\mb{\Co G = \Co(G_1 \times G_2)} on~${\cal R} = {\cal R}_1 \otimes_R
{\cal R}_2.$
\end{theorem}

This reduction is, however, only as abstract computations over vector
spaces.  To~give a concrete quantum circuit for computing the Fourier
transform for the product group, we also need to consider the choices of
bases for the involved transforms in the reduction.  Let~$E_i$ be an
encoding for~$F_i$ and $\mathbf F_i$ the corresponding matrix
representation of~$F_i$, $i=1,2$.  In~ket-notation, the
transform~$\mathbf F'_G$ reads
\begin{equation*}
  \ket{g_1}\ket{g_2} \; \longmapsto \; \big(\mathbf F_1 \ket{g_1}\big)
  \big(\mathbf F_2 \ket{g_2}\big)
\end{equation*}
for all $g_1 \otimes g_2 \in {\cal B}_{\text{time}}^1 \otimes {\cal
B}_{\text{time}}^2$.  

Define the bijections $E_{\text{time}} : {\cal B}_{\text{time}}
\rightarrow \integer_{\eta_1 \eta_2}$ and $E_{\text{freq}} : {\cal
B}_{\text{freq}} \rightarrow \integer_{\eta_1 \eta_2}$~by
\begin{equation}\eqlabel{encoding}
\begin{split}
E_{\text{time}}(\varphi(g_1 \otimes g_2)) &= \eta_2 E_1(g_1)+E_2(g_2)\\
E_{\text{freq}}(\varphi(b_1 \otimes b_2)) &= \eta_2 E_1(b_1)+E_2(b_2).
\end{split}
\end{equation}
Let~$E$ denote the extension of $E_{\text{time}}$ and $E_{\text{freq}}$.
With respect to the encoding~$E$, $F_G$ as defined in
\theoremref{product1} has the matrix representation
\[\mathbf F_G = \mathbf F_1 \otimes_R \mathbf F_2.\]
Applying $\mathbf F_G$ is thus done by applying $\mathbf F_1$ on the
most significant bits, and $\mathbf F_2$ on the least significant bits.
We~have shown
\begin{theorem}\label{thm:product2}
Let $G_1$ and $G_2$ be groups of order $\eta_1$ and $\eta_2$,
respectively.  Suppose we have quantum networks $\mathbf F_1$
and~$\mathbf F_2$ for computing Fourier transforms for $\Co G_1$
and~$\Co G_2$ with respect to the encodings~$E_1$ and~$E_2$,
respectively.  Then the following quantum circuit computes a Fourier
transform for $\Co(G_1 \times G_2)$ with respect to the encoding~$E$ as
defined in~\eq{encoding}.

\begin{center}\leavevmode\hbox{
\setlength{\unitlength}{0.09mm}
\begin{picture}(240,145)(0,55)
\put(70,172){\line(1,0){60}}
\put(70,100){\line(1,0){60}}
\put(164,172){\makebox(0,0){\framebox(68,54){$\mathbf F_1$}}}
\put(164,100){\makebox(0,0){\framebox(68,54){$\mathbf F_2$}}}
\put(198,172){\line(1,0){40}}
\put(198,100){\line(1,0){40}}
\put(42,172){\makebox(0,0){$g_1$}}
\put(42,100){\makebox(0,0){$g_2$}}
\put(100,100){\makebox(0,0){$\scriptstyle /$}}
\put(100,172){\makebox(0,0){$\scriptstyle /$}}
\put(105,140){\makebox(0,0){$\scriptstyle \eta_1$}}
\put(105,68){\makebox(0,0){$\scriptstyle  \eta_2$}}
\end{picture}}
\end{center}
\end{theorem}

As~an application of this, consider the Walsh-Hadamard
transform~\cite{Hadamard1893,Walsh23}, defined as follows.  For any
positive integer~$n$, let
\[\mathbf W_{2^n} \ket{x} = \frac1{\sqrt{2^n}}
   \sum_{y=0}^{2^n-1} (-1)^{\sum_{i=0}^{n-1} x_i y_i} \ket{y},\] 
for each \mb{x=0,\dots,2^n-1,} where $x = x_{n-1} \dots x_0$ and $y =
y_{n-1} \dots y_0$.  This unitary transform can also be defined using
the standard Kronecker product as follows
\begin{equation}
 \mathbf W_2 = \mathbf W, \quad  \mathbf W_{2^{n+1}} = 
 \mathbf W \otimes_R \mathbf W_{2^n}, \; n=1,2,\dotsc
\end{equation}
Appealing to the generalized Kronecker product routine
in~\sect{routines}, we immediately obtain the well-known method for
computing the Walsh-Hadamard transform $\mathbf W_{2^n}$ on a quantum
computer~\cite{DJ92}: apply the transform~$\mathbf W$ on each of the
$n$~qubits.

It~is easy to check that $\mathbf W$ is the Fourier transform for the
cyclic group~$\Bbb Z_2$ of order two.  Thus, by \theoremref{product2},
we have the well-known fact that the Walsh-Hadamard transform coincides
with the Fourier transform for the Abelian group~\mb{\Bbb Z_2^n}.  This
transform has been extensively used in quantum algorithms, for example
by Deutsch and Jozsa~\cite{DJ92}, Simon~\cite{Simon94}, and
Grover~\cite{Grover96,BBHT96}.  One of its advantages is that it can be
computed in only $\Theta(n)$ basic operations~\cite{DJ92}.

Now one might ask if a similar statement holds for subgroups in general.
That is, if $H \subgroup G$ is a subgroup and we encode elements of~$G$
using two registers, the first for coset representatives, the second for
elements from~$H$, to what extent do the gates then need to involve both
registers?  We~consider this question in the next section, and give an
answer to it for some classes of non-Abelian groups.

In~this section, we have carefully distinguished between isomorphic
vector spaces in order to prove \theoremref{product2}.  In~the
following, we will relax slightly upon this to avoid cumbersome
notation.  Let $U$ and~$V$ be any two inner product spaces of dimension
$m$ and~$n$, respectively, with orthonormalized bases
$\{u_1,\dots,u_m\}$ and $\{v_1,\dots,v_n\}$, respectively.  Then the
tensor product $U \otimes V$ and the vector space spanned~by
$\{(u_i,v_j) : 1 \leq i \leq m, 1 \leq j \leq n\}$ are isomorphic under
the natural isomorphism~$\varphi$ given by
$\varphi (u_i \otimes v_j) = (u_i,v_j)$.
When appropriate, we will not distinguish between $u_i \otimes v_j$ and
$(u_i,v_j)$ in the rest of this paper.  Note that the set $\{u_i \otimes
v_j\}$ is an orthonormalized basis for $U \otimes V$.

\section{Adapted representations}\sectlabel{adapted}
In~the previous section, we related a quantum Fourier transform for
$G=G_1 \times G_2$ to quantum Fourier transforms for $G_1$ and~$G_2$.
In~the classical case, relating a Fourier transform of a group to a
Fourier transform to one of its subgroup has shown to be very useful;
see for example~\cite{MR95} and the references therein.  The~main ideas
in this approach are factorization of the group elements, and the use of
an adapted set of representations.  For example, in \sect{product}, we
used the factorization $(g_1,g_2)=(g_1,e_2) \cdot (e_1,g_2)$, where
$e_i$ denotes the identity of~$G_i$, $i=1,2$.

For any subgroup~$H \subgroup G$ and any representation~$\rho$ of~$G$,
let \mb{\rho \downarrow H} denote the representation of~$H$ obtained by
restricting $\rho$ to~$H$.  The representation \mb{\rho \downarrow H} is
unitary but not necessary irreducible.  Recall that we in this paper
assume that all representations are irreducible and unitary.
 
\begin{definition}\label{def:adapted}
Let~$H \subgroup G$ be a subgroup, and $\cal R$ be a complete set of
representations of~$G$.  Then $\cal R$ is called {\em $H$--adapted\/} if
there is a complete set ${\cal R}^H$ of representations of~$H$ such that
the set of restricted representations $({\cal R} \downarrow H) = \{ \rho
\downarrow H : \rho \in {\cal R}\}$ is a set of matrix direct sums of
the representations in~${\cal R}^H$.
\end{definition}

The set~$\cal R$ is said to be {\em adapted to a chain of subgroups\/}
if it is adapted to each subgroup in the chain.  Adapted representations
always exist.

Let \mb{H \subgroup G} be a subgroup of order~$m$, and $T$ a left
transversal for~$H$ in~$G$.  Let~${\cal R}^H$ be a complete set of
representations of~$H$, and let~$\cal R$ be a complete set of
representations of~$G$ that is $H$--adapted relative to~${\cal R}^H$.
Let ${\cal B}^H_{\text{freq}}$ and~${\cal B}_{\text{freq}}$ denote the
collections of normalized matrix coefficients for $H$ and~$G$,
respectively.  Let $\rho \in {\cal R}$ be a representation of
degree~$d$.  The matrix coefficient $\rho_{kl} \in \Co G$ can be written
as a linear sum of the basis-elements ${\cal B}_{\text{time}}$.
\[\rho_{kl} = \sum_{g \in G} \rho_{kl}(g) \,g
  = \sum_{t \in T} \sum_{h \in H} \sum_{i=1}^d
    \rho_{ki}(t) \rho_{il}(h) \,th
  = \sum_{t \in T} \sum_{i=1}^d \rho_{ki}(t) 
    \bigg(\sum_{h \in H} \rho_{il}(h) \,th\bigg).\]

Since~$\cal R$ is $H$--adapted by assumption, $\rho$~is a matrix direct
sum of representations in~${\cal R}^H$.  Therefore, either
\mb{\rho_{il}(h)=0} for all \mb{h \in H}, or there exist $\rho' \in
{\cal R}^H$ of degree~$d'$ and $1 \leq i',l' \leq d'$ such that
\mb{\rho_{il}(h)=\rho'_{i'l'}(h)} for all \mb{h \in H}.  In~the former
case, let \mb{\rho'_{i'l'}} (and \mb{b_{\rho',i',l'}}) denote the zero
vector in~$\Co H$.  Then we have
\[\rho_{kl} = \sum_{t \in T} \sum_{i=1}^d \rho_{ki}(t) 
    \bigg(\sum_{h \in H} \rho'_{i'l'}(h) \,th\bigg),\]
so
\begin{equation}\eqlabel{adapted1}\hspace*{-.4cm}
b_{\rho,k,l} = \sum_{t \in T} \sum_{i=1}^d b_{\rho,k,i}(t) 
    \bigg(\sum_{h \in H} \rho'_{i'l'}(h) \,th\bigg)
  = \sum_{t \in T} \sum_{i=1}^d \sqrt{\frac{m}{d'}} b_{\rho,k,i}(t)
    \bigg(\sum_{h \in H} b_{\rho',i',l'}(h) \,th\bigg).
\end{equation}

A~Fourier transform,~$F_G$, is a change of basis in~$\Co G$ from the
standard basis to a basis of normalized matrix coefficients.  Let~$F_H$
be the Fourier transform for~$\Co H$ on~${\cal R}^H$.  For obtaining an
$H$--adapted method for computing~$F_G$, consider the complex vector
space spanned by the basis
\[T \otimes {\cal B}_{\text{time}}^H = 
  \{t \otimes h \,:\, t \in T, h \in {\cal B}_{\text{time}}^H\}.\]
This vector space is clearly isomorphic to~$\Co G$ under the natural map
$\varphi : \span{T \otimes {\cal B}_{\text{time}}^H} \rightarrow
\span{{\cal B}_{\text{time}}}$ given~by $\varphi(t \otimes h)=th$.
Here, and in the rest of this paper, $\span{\cdot}$ means
$\text{span}(\cdot)$.  Another basis~is
\[{\cal B}_{\text{temp}} = T \otimes {\cal B}_{\text{freq}}^H = 
  \{t \otimes b_{\rho',i',l'} \,:\, t \in T, b_{\rho',i',l'} \in {\cal
  B}_{\text{freq}}^H\},\] 
and using $\varphi$, \eq{adapted1} reads
\[b_{\rho,k,l} = \sum_{t \in T} \sum_{i=1}^d \sqrt{\frac{m}{d'}}
    b_{\rho,k,i}(t) \,\varphi(t \otimes b_{\rho',i',l'}).\]
Let~$V:\span{{\cal B}_{\text{freq}}} \rightarrow \span{{\cal
B}_{\text{temp}}}$ denote the transform
\[V:b_{\rho,k,l} \longmapsto \sum_{t \in T} \sum_{i=1}^d
    \sqrt{\frac{m}{d'}} b_{\rho,k,i}(t) \,t \otimes b_{\rho',i',l'}.\]
By~construction of~$V$,
\[(I \otimes F_H) \circ \varphi^{-1} = V \circ F_G\]
as~illustrated in the following commutative diagram
\[\begin{CD} 
\span{T \otimes {\cal B}_{\text{time}}^H} 
@>\varphi>> \span{{\cal B}_{\text{time}}}\\ 
@V{I \otimes F_H}VV   @VV{F_G}V\\
\span{T \otimes {\cal B}_{\text{freq}}^H} 
@<<V< \span{{\cal B}_{\text{freq}}}
\end{CD}\]

Since~$\varphi$ is an isomorphism, and $F_H$ and~$F_G$ are unitary,
$V$~is invertible.  Let~$U:\span{{\cal B}_{\text{temp}}} \rightarrow
\span{{\cal B}_{\text{freq}}}$ denote the inverse of~$V$, that is,
\begin{equation}\eqlabel{adapted2}
U : \sum_{t \in T} \sum_{i=1}^d \sqrt{\frac{m}{d'}} b_{\rho,k,i}(t)
      \,t \otimes b_{\rho',i',l'} \longmapsto b_{\rho,k,l}
\end{equation}
which maps a vector~$\tilde v \in \span{{\cal B}_{\text{temp}}}$ given
relative to basis ${\cal B}_{\text{temp}}$ to its representation~$\hat v
\in \Co G$ relative to basis ${\cal B}_{\text{freq}}$.  Hence, we have
factorized the Fourier transform~$F_G$ into a product of three unitary
transforms,
\begin{equation}
F_G = U \circ (I \otimes F_H) \circ \varphi^{-1}.
\end{equation}

A~quantum implementation of the adapted method for computing a Fourier
transform can be obtained as follows.  Given a vector $v \in \Co G$, let
$v_t \in \Co G$ denote the vector which is non-zero only on the
coset~$tH$, on which it is given by $v_t(h)=v(th)$ for all $h \in H$.
Initially, we hold the superposition \mb{v = \sum_{g \in {\cal
B}_{\text{time}}} v(g) \ket{g}} and we want to compute the superposition
\mb{\hat v = \sum_{b_i \in {\cal B}_{\text{freq}}} \hat v(b_i)
\ket{b_i}}.  The~quantum routine consists of three steps.  In~the first
step, we apply $\varphi^{-1}$, computing
\[v = \sum_{g \in {\cal B}_{\text{time}}} v(g) \ket{g} \longmapsto 
  \sum_{t \in T} \sum_{h \in \pst{{\cal B}_{\text{time}}}{H}}
  v(th) \ket{t} \ket{h}
  = \sum_{t \in T} \ket{t} \bigg(\sum_{h \in \pst{{\cal
  B}_{\text{time}}}{H}} v_t(h) \ket{h}\bigg).\]
Then, we apply a quantum Fourier transform~$\mathbf F_H$ with respect
to~${\cal R}^H$ to the second register, producing
\[\sum_{t \in T} \ket{t} \bigg(\sum_{b'_i \in \pst{{\cal
  B}_{\text{freq}}}{H}} \hat v_{t}(b'_i) \ket{b'_i} \bigg)
= \sum_{t \in T} \sum_{b'_i \in \pst{{\cal B}_{\text{freq}}}{H}} 
  \hat v_{t}(b'_i) \ket{t} \ket{b'_i} = \tilde v.\]
Finally, in the third step, we apply the linear transform~$\mathbf U$
given by \eq{adapted2}, producing
\[\sum_{b_i \in {\cal B}_{\text{freq}}} \hat v(b_i) \ket{b_i} 
  = \hat v.\]  
The transform~$\mathbf U$ is unitary since the first two steps (that is,
$(I \otimes F_H) \circ \varphi^{-1}$) are unitary and the composition of
the three steps (that is, $F_G$) is unitary.  In~the following sections,
we apply the technique just described to develop quantum Fourier
transforms for some non-Abelian groups.

\subsection{The quaternionic groups}
The {\em quaternionic group\/} $Q_n$ of order~$4n$ is the group
\[Q_n = \langle\: r,c \;:\; r^{2n}=c^4=1,\: cr=r^{2n-1}c,\: c^2=r^n
         \:\rangle.\]
For simplicity, we consider only the case that $n$ is even.  The case
when $n$ is odd is very similar, and in fact slightly simpler.  When $n$
is even, $Q_n$ has a complete set~$\cal R$ consisting of four
one-dimensional and $n-1$ two-dimensional representations
\[\begin{array}{cc}
  \rho^1 \equiv 1 &\quad \rho^2(r) = \rho^2(-c) = 1 \\
  \rho^3(-r) = \rho^3(c) =1 &\quad \rho^4(-r) = \rho^4(-c)=1 
\end{array} \quad
\sigma^i (r) = \begin{bmatrix}\overline \omega^{i}&0\\0&\overline
  \omega^{-i}\end{bmatrix} \quad \sigma^i (c) =
  \begin{bmatrix}0&(-1)^i\\ 1&0\end{bmatrix}\] 
where $1 \leq i < n$ and $\omega=\omega_{2n}$.

The group has a cyclic subgroup~$H$ generated by~$r$ of index two.
Let~$T=\{e,c\}$ be a left transversal for~$H$ in~$Q_n$, and
write~$Q_n=TH$.
Let~${\cal R}^H$ denote the complete set of one-dimensional
representations of~$H$ given in \sect{qft-cyclic}.  The set of
restricted representations of~$\cal R$ is
\[({\cal R} \downarrow H) = \{ \zeta^0, \zeta^n \} \cup
     \{ \zeta^l \oplus \zeta^{2n-l} : 1 \leq l <n \}, \]
so $\cal R$ is $H$--adapted.  Let ${\cal B}_{\text{time}}$, ${\cal
B}_{\text{freq}}$, ${\cal B}_{\text{time}}^H$, and~${\cal
B}_{\text{freq}}^H$ be defined as in \sect{qft}.  Let
\[{\cal B}_{\text{temp}} = T \otimes {\cal B}_{\text{freq}}^H
  = \{ t \otimes b_{\zeta^i,1,1} : t \in T, b_{\zeta^i,1,1} \in {\cal
    B}_{\text{freq}}^H\}.\]

A~main part of the development of subgroup-adapted Fourier transforms is
the determination and implementation of the transform~$U$ defined by
\eq{adapted2}.  For this purpose, consider the matrix coefficient
$\sigma^i_{11} \in \Co Q_n$.  By~writing
\[\sigma^i_{11} = \sum_{t \in T} \sum_{h \in H}
\sigma_{11}^i(th) \,th = \sum_{x \in \integer_{2n}} \overline
\omega^{ix} \,r^x,\]
we have that
\[b_{\sigma^i,1,1} = \frac1{\sqrt{2n}} \sum_{x \in \integer_{2n}}
\overline \omega^{ix} \,r^x = \varphi(e \otimes \zeta^i).\]
Each of the other Fourier coefficients can similarly be written as a
linear sum of the basis-elements ${\cal B}_{\text{temp}}$,
\begin{equation}\eqlabel{transformU:Qn}
\left.\begin{array}{r@{\;}c@{\;}l@{\qquad}r@{\;}c@{\;}l}
b_{\sigma^i,1,1} &=& \varphi(e \otimes \zeta^i) &
b_{\rho^1,1,1} &=& \frac{1}{\sqrt 2} 
(\varphi(e \otimes \zeta^0) + \varphi(c \otimes \zeta^0))\\
b_{\sigma^i,1,2} &=& (-1)^i \, \varphi(c \otimes \zeta^{2n-i})&
b_{\rho^2,1,1} &=& \frac{1}{\sqrt 2} 
(\varphi(e \otimes \zeta^0) - \varphi(c \otimes \zeta^0))\\
b_{\sigma^i,2,1} &=& \varphi(c \otimes \zeta^i)&
b_{\rho^3,1,1} &=& \frac{1}{\sqrt 2} 
(\varphi(e \otimes \zeta^n) + \varphi(c \otimes \zeta^n))\\
b_{\sigma^i,2,2} &=& \varphi(e \otimes \zeta^{2n-i})&
b_{\rho^4,1,1} &=& \frac{1}{\sqrt 2} 
(\varphi(e \otimes \zeta^n) - \varphi(c \otimes \zeta^n))\\
\end{array}\right\}
\end{equation}
where $i=1,\dots,n-1$.

\eq{transformU:Qn} defines the transform~$U: \span{{\cal
B}_{\text{temp}}} \rightarrow \span{{\cal B}_{\text{freq}}}$ appearing
in the factorization $F_G = U \circ (I \otimes F_H) \circ \varphi^{-1}$.
What remains in order to obtain a concrete circuit computing the Fourier
transform~$F_G$, is an encoding of the bases for the transforms.  With
respect to the encoding
\[E_{\text{time}}^H(r^k) = k \qquad E_{\text{freq}}^H(\zeta^i) = i,\] 
the Fourier transform~$F_H$ for~$\Co H$ defined by \eq{dft} has the
matrix representation~$\mathbf F_H$.  Let therefore the encoding of
${\cal B}_{\text{time}}$ be given by $E_{\text{time}}(c^j r^k) = 2nj +
k$, and the encoding of ${\cal B}_{\text{temp}}$ be given by
$E_{\text{temp}}(c^j \otimes \zeta^i) = 2nj + i$.  With respect to this
encoding, the transform $(I \otimes F_H) \circ \varphi^{-1}$ has the
matrix representation $\mathbf I_2 \otimes_R \mathbf F_H$.

Computing the $U$ transform with respect to $E_{\text{temp}}$ is very
simple
\[\ket{j}\ket{i} \longmapsto
\begin{cases}
\;\frac1{\sqrt 2}\big(\ket{0} + (-1)^j \ket{1}\big) \ket{i}
 &\text{ if $i=0$ or $i=n$}\\
\;(-1)^j \ket{j}\ket{i} &\text{ if $i>n$ and $i$ is odd}\\
\;\ket{j}\ket{i} &\text{ otherwise.}
\end{cases}\]
So, given a network computing the Fourier transform $\mathbf F = \mathbf
F_{2n}$ for the cyclic group of order~$2n$, a network computing the
Fourier transform for the quaternionic group~$Q_n$ can be constructed as
follows.

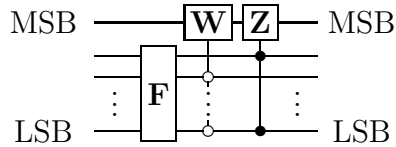
\begin{center}\leavevmode\hbox{
\setlength{\unitlength}{0.09mm}
\begin{picture}(600,200)(0,140)
\put(80,150){\line(1,0){70}}
\put(80,230){\line(1,0){70}}
\put(80,260){\line(1,0){70}}
\put(410,150){\line(-1,0){154.5}}
\put(410,230){\line(-1,0){154.5}}
\put(410,260){\line(-1,0){210}}
\put(200,150){\line(1,0){40.5}}
\put(200,230){\line(1,0){40.5}}
\put(80,310){\line(1,0){134}}
\put(282,310){\line(1,0){18}}
\put(350,310){\line(1,0){60}}
\put(110,201){\makebox(0,0){$\vdots$}}
\put(380,201){\makebox(0,0){$\vdots$}}
\put(175,205){\makebox(0,0){\framebox(50,140){$\mathbf F$}}}
\put(248,310){\makebox(0,0){\framebox(68,50){$\mathbf W$}}}
\put(248,150){\circle{15}}
\put(248,230){\circle{15}}
\put(248,157.5){\line(0,1){7}}
\put(248.3,201){\makebox(0,0){$\vdots$}}
\put(248,222.5){\line(0,-1){7}}
\put(248,285){\line(0,-1){47.5}}
\put(325,310){\makebox(0,0){\framebox(50,50){$\mathbf Z$}}}
\put(325,260){\circle*{15}}
\put(325,150){\circle*{15}}
\put(325,150){\line(0,1){135}}
\put(5,150){\makebox(0,0){LSB}}
\put(5,310){\makebox(0,0){MSB}}
\put(475,310){\makebox(0,0){MSB}}
\put(475,150){\makebox(0,0){LSB}}
\end{picture}}
\end{center}

Note that $\mathbf Z$ and $\mathbf W$, both defined in \sect{routines},
operate on distinct states and thus commute.  If~we, on the circuit
given above, remove the $\mathbf Z$~gate, then we have a circuit that
computes a Fourier transform for both the dihedral and the semidihedral
group of order~$4n$.  In~the following, we show that this is no
coincidence.

\subsection{Metacyclic Groups}
In~this section, we give a general quantum circuit for computing a
Fourier transform for a class of metacyclic groups.  A~group is called
{\em metacyclic\/} if it contains a cyclic normal subgroup~$H$ so that
the quotient group~$G/H$ is also cyclic.  Let~$G = \{b^j a^i : 0 \leq j
< q,\: 0 \leq i < m\}$ be a metacyclic group where
\[b^{-1}ab=a^r, \quad b^q=a^s, \quad a^m=1\]
and with $(m,r)=1,\: m|s(r-1), \text{ and~$q$ prime}$.  Let~$d=(r-1,m)$.
The group has a cyclic subgroup~$H$ generated by~$a$ of index~$q$.
Let~$T=\{b^j: 0 \leq j < q\}$ be a left transversal for~$H$ in~$G$, and
write~$G=TH$.  Let $E_{\text{time}}: {\cal B}_{\text{time}} \rightarrow
\integer_{qm}$ be the encoding of ${\cal B}_{\text{time}}$ given by
\[b^j a^i \longmapsto mj+i.\]

\begin{theorem}\label{thm:meta}
The following network computes a Fourier transform for~$\Co G$ up to
phase factors with respect to~$E_{\text{time}}$.  Here $\omega =
\omega_{qd}^s$.

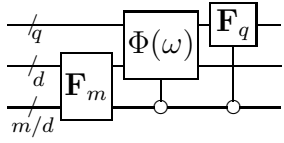
\begin{center}
\leavevmode
\hbox{
\setlength{\unitlength}{0.11mm}
\begin{picture}(330,145)(70,85)
\put(70,100){\line(1,0){65}}
\put(70,150){\line(1,0){65}}
\put(70,200){\line(1,0){141}}
\put(211,150){\line(-1,0){16}}
\put(299,150){\line(1,0){101}}
\put(299,200){\line(1,0){16}}
\put(370,200){\line(1,0){30}}
\put(255,100){\circle{15}}
\put(342.5,100){\circle{15}}
\put(195,100){\line(1,0){52.5}}
\put(335,100){\line(-1,0){72.5}}
\put(350,100){\line(1,0){50}}
\put(255,107.5){\line(0,1){27.5}}
\put(342.5,107.5){\line(0,1){67.5}}
\put(100,200){\makebox(0,0){$\scriptstyle /$}}
\put(100,150){\makebox(0,0){$\scriptstyle /$}}
\put(100,100){\makebox(0,0){$\scriptstyle /$}}
\put(105,186){\makebox(0,0){$\scriptstyle q$}}
\put(107,135){\makebox(0,0){$\scriptstyle d$}}
\put(100,77){\makebox(0,0){$\scriptstyle m/d$}}
\put(165,125){\makebox(0,0){\framebox(60,80){$\mathbf F_m$}}}
\put(255,175){\makebox(0,0){\framebox(88,80){$\Phi(\omega)$}}}
\put(342.5,200){\makebox(0,0){\framebox(55,50){$\mathbf F_q$}}}
\end{picture}}
\end{center}
\end{theorem}

The phase factors involved in the theorem depend on the actual group
structure.  Before proving the theorem, we consider the representations
of the group.  The group~$G$ has $qd$~one-dimensional representations,
\mb{\{\rho^{ij}\}_{i,j=0}^{i=d-1,j=q-1}}, each given~by
\[\rho^{ij}(a) = \overline \omega_d^i \qquad 
  \rho^{ij}(b) = \overline \omega_q^j \overline \omega_{qd}^{is}.\] 
Let ${\cal R}^H$ be the complete set of representations of~$H$ given in
\sect{qft-cyclic}.  For every $\zeta^i \in {\cal R}^H$ define the {\em
induced\/} representation $\bar \zeta^i: G \rightarrow
\text{GL}_q(\Co)$~by
\[a \longmapsto \begin{bmatrix}\overline \omega_m^i&&\\ &\ddots&\\
                 &&\overline \omega_m^{i r^{q-1}}\end{bmatrix} \qquad
  b \longmapsto \begin{bmatrix}&&&\overline \omega_m^{is} \\ 1&&&\\ 
                 &\ddots&& \\ &&1&\end{bmatrix}.\]
The group~$G$ has an $H$--adapted set of representations ${\cal R}$
consisting of the $qd$~one-dimensional and \mb{(m-d)/q}
\,$q$--dimensional representations.  The $q$--dimensional
representations in~${\cal R}$ are all induced
representations~\cite{CR62}.

The matrix coefficient $\rho^{ij} \in \Co G$ can be written as a linear
sum of the basis-elements ${\cal B}_{\text{temp}} = T \otimes {\cal
B}_{\text{freq}}^H$,
\begin{align*}
\rho^{ij} &= \sum_{g \in G} \rho^{ij}(g) \,g\\
&= \sum_{k \in \Bbb Z_q} \sum_{x \in \Bbb Z_m} 
   \rho^{ij}(b^k a^x) \,b^k a^x\\
&= \sum_{k \in \Bbb Z_q} \rho^{ij}(b^k) \sum_{x \in \Bbb Z_m} 
   \rho^{ij}(a^x) \,b^k a^x\\
&= \sum_{k \in \Bbb Z_q} \overline \omega_q^{jk} 
   \big(\overline \omega_{qd}^{sik}
   \sum_{x \in \Bbb Z_m} \overline \omega_m^{xim/d} 
   \,b^k a^x\big)\\
&= \sqrt m \sum_{k \in \Bbb Z_q} \overline \omega_q^{jk} 
   \big(\overline \omega_{qd}^{sik}
   \,\varphi(b^k \otimes b_{\zeta^{im/d},1,1})\big)
\end{align*}
so, by definition of $U:\span{{\cal B}_{\text{temp}}} \rightarrow
\span{{\cal B}_{\text{freq}}}$ as given in \sect{adapted},
\begin{equation}\eqlabel{one-dim} 
U^{-1}: b_{\rho^{ij},1,1} \longmapsto \frac1{\sqrt q} \sum_{k \in \Bbb
Z_q} \overline \omega_q^{jk} \big(\overline \omega_{qd}^{sik} \,b^k
\otimes b_{\zeta^{im/d},1,1}\big).
\end{equation}

We~refer to a matrix coefficient of an induced representation as an {\em
induced\/} matrix coefficient.  Any induced matrix coefficient $\bar
\zeta^i_{kl} \in \Co G$ is non-zero on exactly one coset of~$H$.  For
example, $\bar \zeta^i_{kl} = {\bar \zeta^i}_{31}$ is non-zero on the
coset $b^2 H$.  In~general, the matrix coefficient $\bar \zeta^i_{kl}
\in \Co G$ can be written as a linear sum of the basis-elements~${\cal
B}_{\text{temp}}$ as follows.
\begin{align*}
\bar \zeta^i_{kl} 
&= \sum_{g \in G} \bar \zeta^i_{kl}(g) \,g\notag\\
&= \sum_{t \in T} \sum_{h \in H} \bar \zeta^i_{kl}(t h) \,t h\notag\\
&= \sum_{t \in T} \bar \zeta^i_{kl}(t) \sum_{h \in H} 
      \bar \zeta^i_{ll}(h)  \,t h\notag\\
&= \bar \zeta^i_{kl}(b^{k-l}) \sum_{h \in H} 
      \bar \zeta^i_{ll}(h)  \,b^{k-l} h\notag\\
&= \bar \zeta^i_{kl}(b^{k-l}) \sum_{x \in \integer_m} 
      \overline \omega_m^{ir^lx}  \,b^{k-l} a^x\notag\\
&= \sqrt m \,\phi\, \varphi(b^{k-l} \otimes b_{\zeta^{ir^l},1,1}).
\end{align*}
Here, $\phi = \bar \zeta^i_{kl}(b^{k-l})$ is some \th{m} root of unity.
So 
\begin{equation}\eqlabel{coef}
U^{-1}: b_{\bar \zeta^i,k,l} \longmapsto \phi \,b^{k-l} \otimes
b_{\zeta^{ir^l},1,1}.
\end{equation}
that is, $b_{\bar \zeta^i,k,l} \in {\cal B}_{\text{freq}}$ is mapped
by~$U^{-1}$ to one of the basis-elements ${\cal B}_{\text{temp}}$ up to
a phase-factor.  To~find an expression for~$U$ instead of $U^{-1}$, we
need \lemmaref{reduc} which easily follows from the following lemma for
which a proof can be found, for example, in~\cite[Lemma (47.8)]{CR62}.

\begin{lemma}\label{lm:CR} 
The induced representation $\bar \zeta^i$ is reducible if and only if
there exists a $j$, \mb{1 \leq j \leq q-1}, such that $i r^j \equiv i
\pmod{m}$.
\end{lemma}

\begin{lemma}\label{lm:reduc}
Let~${\bar \zeta^i}_{kl}$ be any induced matrix coefficient.  If~$\bar
\zeta^i$ is irreducible, then $ir^l$ is a not multiple of~$m/d$.
\end{lemma}

\begin{proof}
To~prove the contrapositive, suppose that $ir^l \equiv 0 \pmod{m/d}$.
Since $(r,m)=1$, $i \equiv 0 \pmod{m/d}$, so $id \equiv 0 \pmod{m}$.
Since $d$ divides~$r-1$, we have $ir \equiv i \pmod{m}$ and the
statement follows from \lemmaref{CR}.
\end{proof}

\begin{lemma}\label{lm:Utransform}
The transform~$U:\span{{\cal B}_{\text{temp}}} \rightarrow \span{{\cal
B}_{\text{freq}}}$ is given by
\[U(b^k \otimes b_{\zeta^{x},1,1}) = \begin{cases} 
\phi \,b_i & \text{ if $x$ is not a multiple of $m/d$}\\ \frac1{\sqrt q}
\omega_{qd}^{sik} \sum_{j \in \integer_q} \omega_q^{jk}
\,b_{\rho^{ij},1,1} & \text{ if $x=im/d$}.
\end{cases}\]
Here, $\phi$ is some \th{m} root of unity and $b_i \in {\cal
B}_{\text{freq}}$, both depending on the value of $k$ and~$x$.
\end{lemma}

\begin{proof}
Write ${\cal B}_{\text{temp}}$ as a disjoint union of two sets, ${\cal
B}^1_{\text{temp}}$ and~${\cal B}^2_{\text{temp}}$, where ${\cal
B}^1_{\text{temp}} = T \otimes \{b_{\zeta^x,1,1} : \text{ $x$ is a
multiple of~$m/d$}\}$.  Write similarly ${\cal B}_{\text{freq}}$ as a
disjoint union of two sets, ${\cal B}^1_{\text{freq}}$ and~${\cal
B}^2_{\text{freq}}$, where ${\cal B}^1_{\text{freq}} =
\{b_{\rho^{ij},1,1} : 0 \leq i < d, 0 \leq j < q\}$.  We~first show that
\begin{equation}\eqlabel{span}
\span{U^{-1}({\cal B}^2_{\text{freq}})} = \span{{\cal
B}^2_{\text{temp}}}
\end{equation}
by a simple counting argument.  For each of the $q(m-d)$ elements
$b_{\bar \zeta^i,k,l} \in {\cal B}^2_{\text{freq}}$, we have that
$\zeta^i \in {\cal R}$ is irreducible.  By~\lemmaref{reduc}, $ir^l$ is
not a multiple of~$m/d$, and therefore $U^{-1}(b_{\bar \zeta^i,k,l}) \in
\span{{\cal B}^2_{\text{temp}}}$.  Since ${\cal B}^2_{\text{freq}}$ and
${\cal B}^2_{\text{temp}}$ has the same cardinality, and since~$U$ is
unitary, \eq{span} follows.  By~\eq{coef}, the first case in the lemma
follows .

By~the unitarity of~$U$, we also have that 
\[\span{U^{-1}({\cal B}^1_{\text{freq}})} = \span{{\cal
B}^1_{\text{temp}}}.\] 
The action of~$U^{-1}$ on ${\cal B}^1_{\text{freq}}$ is given by
\eq{one-dim}, and the action of its inverse (that is, of~$U$) on ${\cal
B}^1_{\text{temp}}$ is given by the second case in the lemma.
\end{proof}

Let~$U_1:\span{{\cal B}_{\text{temp}}} \rightarrow \span{{\cal
B}_{\text{freq}}}$ denote the unitary transform which acts on~${\cal
B}^1_{\text{temp}}$ as~$U$, and which on~${\cal B}^2_{\text{temp}}$ is
given by $b^{k-l} \otimes b_{\zeta^{ir^l},1,1} \longmapsto \,b_{\bar
\zeta^i,k,l}$.  Since we are only interested in a quantum network that
computes a Fourier transform for~$\Co G$ up to phase factors, by
\lemmaref{Utransform}, it suffices to implement~$U_1$ instead of~$U$.
In~conclusion, we have shown that the transform
\[F_G = U_1 \circ (I \otimes F_m) \circ \varphi^{-1}\]
is the Fourier transform for~$\Co G$ on~${\cal R}$ up to phase factors.
Here, $F_m = F_H$~is the Fourier transform for $\Co H$ defined in
\sect{qft-cyclic}.

We~now consider the implementation of~$F_G$.  The encoding
$E_{\text{time}}: {\cal B}_{\text{time}} \rightarrow \integer_{qm}$ is
given above.  Let $E_{\text{temp}}: {\cal B}_{\text{temp}} \rightarrow
\integer_{qm}$ be given by $b^j \otimes \zeta^i \mapsto mj+i$.  With
respect to $E_{\text{time}}$ and~$E_{\text{temp}}$, the transform $(I
\otimes F_H) \circ \varphi^{-1}$ is implemented by $\mathbf I_q
\otimes_R \mathbf F_m$.  Let the encoding $E_{\text{freq}}$ of the
matrix coefficients arising from the one-dimensional representations be
given by $\rho^{ij} \mapsto jm+im/d$.  

With respect to $E_{\text{temp}}$ and $E_{\text{freq}}$, the
transform~$U_1$ can be represented~by
\[\ket{km+im/d+x} \longmapsto \begin{cases}
\ket{km+im/d+x} & \text{ if $1<x<m/d$}\\ \omega_{qd}^{sik} \frac1{\sqrt
q} \sum_{j \in \integer_q} \omega_q^{jk} \,\ket{jm+im/d+x} & \text{ if
$x=0$.}\end{cases}\]
Here, $k \in \integer_q, i \in \integer_d, \text{ and } x \in
\integer_{m/d}$.

Written as a generalized Kronecker product, this~is
\[\big((\mathbf F_q \otimes_R \mathbf I_d) \times
  \Phi_{qd}(\omega_{qd}^s), \mathbf I_{qd}, \dots, \mathbf I_{qd}\big)
  \otimes_R \mathbf I_{m/d}.\] 
Thus, with respect to $E_{\text{time}}$ and $E_{\text{freq}}$, $F_G$ is
computed up to phase factors by a quantum circuit implementing
\[\mathbf F_G^\phi = \big(\big((\mathbf F_q \otimes_R \mathbf I_d)
  \times \Phi_{qd}(\omega_{qd}^s), \mathbf I_{qd}, \dots, \mathbf
  I_{qd}\big) \otimes_R \mathbf I_{m/d}\big) \times \big(\mathbf I_q
  \otimes_R \mathbf F_m\big).\]
\theoremref{meta} follows.

\section{Fourier transforms related to error-correction}
\sectlabel{errorcorr}
In~this section, we give a quantum circuit for computing a Fourier
transform for a certain subgroup~$E_n$ of the orthogonal
group~$\text{O}(2^n) = \{\mathbf A \in \text{GL}_{2^n}(\Co) : \mathbf A
\mathbf A^t = \mathbf I\,\}$.  The group~$E_n$ was used independently by
Gottesman~\cite{Gottesman96} and Calderbank {\it et.~al.}~\cite{CRSS97}
to give a group theoretical framework for studying quantum
error-correcting codes.  

For all $i=1,\dots,n$, define
\[\begin{array}{c@{\;= \mathbf I_{2^{i-1}}\, \otimes_R\;}c%
@{\;\otimes_R \;\mathbf I_{2^{n-i}}}l}
\mathbf X_i & \mathbf X &\\
\mathbf Z_i & \mathbf Z &\\
\mathbf Y_i & \mathbf Y &,\\
\end{array}\]
where $\mathbf X, \mathbf Z, \text{ and } \mathbf Y$ are given as
in~\sect{routines}.  The group~$E_n$ is the group generated by
these~$3n$ unitary matrices.  Its order is~$2 \cdot 4^n$.  Every element
squares to either $\mathbf I$ or $-\mathbf I$, and two elements either
commute or anti-commute.  When $n=0$, $E_n = \{[\pm 1]\}$ is a cyclic
group of order two, and if~$n=1$, $E_n$ is isomorphic to~$D_4$.  For
larger~$n$, $E_n$ is isomorphic to $D_4^n / K_n$ where $K_n$ is a normal
subgroup isomorphic to~$\integer_2^{n-1}$.  Given $a,c \in
\integer_2^n$, $a=(a_1,\dots,a_n)$ and $c=(c_1,\dots,c_n)$, let $\mathbf
X(a)$ and $\mathbf Z(c)$, respectively, denote the elements
$\prod_{i=1}^n \mathbf X_i^{a_i}$ and $\prod_{i=1}^n \mathbf Z_i^{c_i}$,
respectively.  Then every element $g$ of~$E_n$ can be written uniquely
in the form
\begin{equation}\eqlabel{gform}
g = (-\mathbf I)^\lambda \mathbf X(a) \mathbf Z(c)
\end{equation}
where $\lambda \in \integer_2$, and $a,c \in \integer_2^n$.  We~denote
$g$ by the 3--tuple $(\lambda,a,c)$.  By~rewriting \eq{gform}, $g$ can
be written as a right Kronecker product
\begin{equation}\eqlabel{recform:sigma}
g = (\lambda,a,c) = \big((- \mathbf I_2)^\lambda \mathbf X^{a_1} \mathbf
Z^{c_1}\big) \otimes_R \big(\mathbf X^{a_2} \mathbf Z^{c_2}\big)
\otimes_R \dots \otimes_R \big(\mathbf X^{a_n} \mathbf Z^{c_n}\big).
\end{equation}

For $n \geq 1$, let $H \subgroup E_n$ be the subgroup $\{(\lambda,a,c)
\in E_n : a_n = c_n = 0\}$ of index~$4$, and identify $E_{n-1}$ with~$H$
in~$E_n$.  Write $E_n = T E_{n-1}$ where $T=\{\mathbf X_n^{a_n} \mathbf
Z_n^{c_n} : a_n, c_n \in \integer_2\}$ is a left transversal
for~$E_{n-1}$ in~$E_n$.  The group~$E_n$ has a complete set~${\cal
R}_{(n)}$ of $1+2^{2n}$ inequivalent, irreducible and unitary
representations, all but one of dimension one (except for $n=0$ where
both representations, denoted $\sis{0}{}{}\rho$ and~$\sis{0}{}{}\sigma$,
are one-dimensional).  The $2^{2n}$ one-dimensional representations
$\{\sis{n}{}{xz}\rho\}_{x,z \in \integer_2^n}$ are given~by
\[\sis{n}{}{xz}\rho (g) = \sis{n}{}{xz}\rho((\lambda,a,c))
  = (-1)^{{x \cdot a} + {z \cdot c}}.\] 
The last representation,~$\sis{n}{}{}\sigma$, has dimension~$2^n$ and is
the group itself.  From \eq{recform:sigma}, we have the following
recursive expression for the \th{(kk_n,ll_n)} entry of the element
$g=(\lambda,aa_n,cc_n) \in E_n$, $a,c,k,l \in \integer_2^{n-1}$,
\[\sis{n}{kk_nll_n}{}\sigma ((\lambda,aa_n,cc_n)) = (-1)^{l_nc_n} 
  \,\delta_{d_na_n} \sis{n-1}{kl}{}\sigma ((\lambda,a,c))\] 
where $d_n = k_n \oplus l_n \in \integer_2$.  Hence, ${\cal R}_{(n)}$ is
$E_{n-1}$--adapted relative to~${\cal R}_{(n-1)}$.

We~use the concept of adapted representations to find a Fourier
transform for~$\Co E_n$.  Let the bases ${\cal B}_{\text{time}}$, ${\cal
B}_{\text{freq}}$, ${\cal B}^H_{\text{time}}$, ${\cal
B}^H_{\text{freq}}$, and ${\cal B}_{\text{temp}} = T \otimes {\cal
B}^H_{\text{freq}}$ be given as in \sect{qft}.  Let $\varphi : \span{T
\otimes {\cal B}^H_{\text{time}}} \rightarrow \span{{\cal
B}_{\text{time}}}$ denote the natural isomorphism defined in
\sect{adapted}.

The matrix coefficients of~${\cal R}_{(n)}$ can be written as linear
sums of the basis-elements~${\cal B}_{\text{temp}}$
\begin{align*}
\sis{n}{}{xx_nzz_n}\rho &= \sum_{\lambda \in \integer_2} 
\sum_{a,c \in \integer_2^n} 
\sis{n}{}{xx_nzz_n}\rho((\lambda,a,c)) \,(\lambda,a,c)\\
&= \sum_{a_n \in \integer_2} \sum_{c_n \in \integer_2} 
(-1)^{a_nx_n+c_nz_n} \,\varphi(\mathbf X_n^{a_n} \mathbf Z_n^{c_n} 
\otimes \sis{n-1}{}{xz}\rho)\\
\sis{n}{kk_nll_n}{}\sigma &= \sum_{\lambda \in \integer_2}
\sum_{a,c \in \integer_2^n} 
\sis{n}{kk_nll_n}{}\sigma((\lambda,a,c)) \,(\lambda,a,c)\\
&= \sum_{c_n \in \integer_2} (-1)^{c_n l_n} \bigg( 
\sum_{\lambda \in \integer_2} \sum_{\pst{a,c \in \integer_2}{n-1}}
\sis{n-1}{kl}{}\sigma((\lambda,a,c)) \,(\lambda,ad_n,cc_n) \bigg)\\
&= \sum_{c_n \in \integer_2} 
(-1)^{c_n l_n} \,\varphi(\mathbf X_n^{d_n}
\mathbf Z_n^{c_n} \otimes \sis{n-1}{kl}{}\sigma)
\end{align*}
where $x,z,k,l \in \integer_2^{n-1}$ and $d_n = k_n \oplus l_n \in
\integer_2$.  Hence
\begin{equation}\eqlabel{transformU:En}\left.
\begin{split}
b_{\sis{n}{}{xx_nzz_n}\rho,1,1} &= \frac12 \sum_{a_n \in \integer_2}
\sum_{c_n \in \integer_2} (-1)^{a_nx_n+c_nz_n} \,\varphi(\mathbf
X_n^{a_n} \mathbf Z_n^{c_n} \otimes b_{\sis{n-1}{}{xz}\rho,1,1})\\
b_{\sis{n}{}{}\sigma,kk_n,ll_n} &= \frac1{\sqrt 2}
\sum_{c_n \in \integer_2} (-1)^{c_nl_n} \,\varphi(\mathbf X_n^{a_n}
\mathbf Z_n^{c_n} \otimes b_{\sis{n-1}{}{}\sigma,k,l})
\end{split}\:\right\}
\end{equation}
where $k_n=a_n \oplus l_n \in \integer_2$.

\eq{transformU:En} seems to have the form of two $\mathbf W$ transforms
for the one-dimensional representations~$\rho$, and a single $\mathbf W$
transform for the $\sigma$~representation.  With respect to an
appropriate encoding, this is indeed the case.  Choose the encoding
$E^{(n)}: E_n \rightarrow \integer_2^{2n+1}$, $n \geq 0$,
\begin{align*}
E^{(0)}_{\text{time}}((\lambda,\epsilon,\epsilon)) &= \lambda\\
E^{(n)}_{\text{time}}((\lambda,aa_n,cc_n)) &=
 E^{(n-1)}_{\text{time}}((\lambda,a,c)) a_n c_n\\
E^{(n)}_{\text{temp}}(\mathbf X_n^{a_n} \mathbf Z_n^{c_n} \otimes
 \sis{n-1}{kl}{}\sigma) &=
 E^{(n-1)}_{\text{freq}}(\sis{n-1}{kl}{}\sigma) a_n c_n\\ 
E^{(0)}_{\text{freq}}(\sis{0}{}{}\rho) &= 0\\
E^{(0)}_{\text{freq}}(\sis{0}{}{}\sigma) &= 1\\
E^{(n)}_{\text{freq}}(\sis{n}{}{xx_nzz_n}\rho) &= 
 E^{(n-1)}_{\text{freq}}(\sis{n-1}{}{xz}\rho) x_n z_n \\
E^{(n)}_{\text{freq}}(\sis{n}{kk_nll_n}{}\sigma) &= 
 E^{(n-1)}_{\text{freq}}(\sis{n-1}{kl}{}\sigma) a'_n l_n
  \quad \text{(where $a'_n = k_n \oplus l_n$).}
\end{align*}
On~the right hand side of the expressions, think of the images of the
encoding as binary strings with standard string concatenation.

With respect to this encoding, the transform~$U:\span{{\cal
B}_{\text{temp}}} \rightarrow \span{{\cal B}_{\text{freq}}}$, for $n
\geq 1$, can be represented~by
\[\ket{\lambda s a_n c_n} \longmapsto \begin{cases} 
\frac12 \sum_{x_n \in \integer_2} \sum_{z_n \in \integer_2}
(-1)^{a_nx_n+c_nz_n} \ket{\lambda s x_n z_n} &\text{ if $\lambda=0$}\\ 
\frac1{\sqrt 2} \sum_{l_n \in \integer_2}
(-1)^{c_nl_n} \ket{\lambda s a_n l_n} &\text{ if $\lambda=1$}
\end{cases}\]
where $\lambda \in \integer_2, s \in \integer_2^{2n-2}$ and $a_n, c_n
\in \integer_2$.  As~a generalized Kronecker product, this reads 
\begin{equation}\eqlabel{transformU:En:kro}
\big(\mathbf I_2 \otimes_R (\mathbf I_{2^{2n-2}} \otimes_R 
  \mathbf W, \mathbf I_{2^{2n-1}})\big) \otimes_R \mathbf W.
\end{equation}

For $n \geq 1$, let $\mathbf E$ be a quantum circuit computing the
Fourier transform for $\Co E_{n-1}$ on ${\cal R}_{(n-1)}$ with respect
to the above encoding.  Then, by \eq{transformU:En:kro}, the following
network computes the Fourier transform for $\Co E_{n}$ on ${\cal
R}_{(n)}$, also with respect to the above encoding.

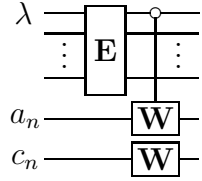
\begin{center}\leavevmode\hbox{
\setlength{\unitlength}{0.09mm}
\begin{picture}(300,240)(10,90)
\put(70,316){\line(1,0){61}}
\put(70,286){\line(1,0){61}}
\put(100,265){\makebox(0,0){$\vdots$}}
\put(268,265){\makebox(0,0){$\vdots$}}
\put(70,220){\line(1,0){61}}
\put(70,160){\line(1,0){130}}
\put(70,100){\line(1,0){130}}
\put(160,261){\makebox(0,0){\framebox(58,130){$\mathbf E$}}}
\put(189,286){\line(1,0){109}}
\put(189,220){\line(1,0){109}}
\put(234,184){\line(0,1){125.5}}
\put(234,316){\circle{15}}
\put(241.5,316){\line(1,0){56.5}}
\put(189,316){\line(1,0){37.5}}
\put(234,160){\makebox(0,0){\framebox(68,48){$\mathbf W$}}}
\put(234,100){\makebox(0,0){\framebox(68,48){$\mathbf W$}}}
\put(268,160){\line(1,0){30}}
\put(268,100){\line(1,0){30}}
\put(39,316){\makebox(0,0){$\lambda$}}
\put(42,160){\makebox(0,0){$a_n$}}
\put(42,100){\makebox(0,0){$c_n$}}
\end{picture}}
\end{center}

For $n=0$, the one-bit network consisting only of the $\mathbf W$ 
transform computes the Fourier transform.  Thus, expanding this
recursively defined network given above, we have

\begin{theorem}
The following network computes a Fourier transform for~$\Co E_n$.
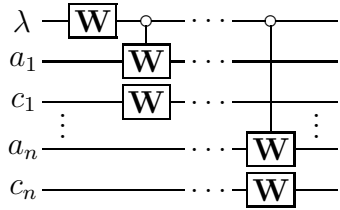
\begin{center}\leavevmode\hbox{
\setlength{\unitlength}{0.09mm}
\begin{picture}(400,280)(30,90)
\put(90,350){\line(1,0){40}}
\put(90,290){\line(1,0){119}}
\put(90,230){\line(1,0){119}}
\put(164,350){\makebox(0,0){\framebox(68,48){$\mathbf W$}}}
\put(243,290){\makebox(0,0){\framebox(68,48){$\mathbf W$}}}
\put(243,230){\makebox(0,0){\framebox(68,48){$\mathbf W$}}}
\put(427,160){\makebox(0,0){\framebox(68,48){$\mathbf W$}}}
\put(427,100){\makebox(0,0){\framebox(68,48){$\mathbf W$}}}
\put(243,350){\circle{15}}
\put(427,350){\circle{15}}
\put(335,350){\makebox(0,0){$\hdots$}}
\put(335,290){\makebox(0,0){$\hdots$}}
\put(335,230){\makebox(0,0){$\hdots$}}
\put(335,160){\makebox(0,0){$\hdots$}}
\put(335,100){\makebox(0,0){$\hdots$}}
\put(90,100){\line(1,0){210}}
\put(90,160){\line(1,0){210}}
\put(277,230){\line(1,0){23}}
\put(277,290){\line(1,0){23}}
\put(250.5,350){\line(1,0){49.5}}
\put(393,100){\line(-1,0){23}}
\put(393,160){\line(-1,0){23}}
\put(525,230){\line(-1,0){155}}
\put(525,290){\line(-1,0){155}}
\put(419.5,350){\line(-1,0){49.5}}
\put(120,207){\makebox(0,0){$\vdots$}}
\put(495,207){\makebox(0,0){$\vdots$}}
\put(198,350){\line(1,0){37.5}}
\put(525,160){\line(-1,0){64}}
\put(525,100){\line(-1,0){64}}
\put(243,342.5){\line(0,-1){28.5}}
\put(427,342.5){\line(0,-1){158.5}}
\put(525,350){\line(-1,0){90.5}}
\put(59,350){\makebox(0,0){$\lambda$}}
\put(62,100){\makebox(0,0){$c_n$}}
\put(62,160){\makebox(0,0){$a_n$}}
\put(62,230){\makebox(0,0){$c_1$}}
\put(62,290){\makebox(0,0){$a_1$}}
\end{picture}}
\end{center}
\end{theorem}

\section{Conclusion}
The problem of finding efficient quantum algorithms computing a given
unitary transform can be formulated as a purely matrix factorization
problem.  Let~$\cal U$ be a set of basic unitary matrices.  Given a
unitary matrix~$U$ of dimension~$(n \times n)$, can~$U$ be factorized
into a product of basic unitary matrices such that the number of
components in this product is polynomial bounded in $\log(n)$?
Previously, the only operations considered allowed in this product have
been the basic binary matrix operations: multiplication and standard
Kronecker product.  In~this paper, we have shown that allowing a
generalization of the latter, efficient networks can still be obtained.

This generalized operation has several advantages.  First of all, it
gives a new tool when searching for factorizations of unitary matrices.
The two new quantum networks given in \sect{qwt} implementing the
wavelet transforms were found this way.  Secondly, it gives a nice
compact mathematical description of more complex transforms.  Thirdly,
it directly gives quantum networks for computing unitary transforms
which already were known to be expressible by generalized Kronecker
products.  This is for example the case for the Fourier transforms for
the finite Abelian groups.

In~this paper, we have also discussed the issue of computing Fourier
transforms for finite non-Abelian groups.  We~have given a definition of
such computations on quantum computers, and especially we have given a
slightly relaxed definition where we only compute a Fourier transform up
to phase factors.  Using this latter definition, we have devised a
quantum network computing a Fourier transform for a class of meta-cyclic
groups---even without completely knowing the group structure.  This
relaxed definition is in particular useful if the computation are to be
followed by a measurement~\cite{GN95}, as for example in the algorithms
of Deutsch and Jozsa~\cite{DJ92}, Simon~\cite{Simon94},
Shor~\cite{Shor97}, and Boneh and Lipton~\cite{BL95}.

We~have also given a simple quantum circuit computing a Fourier
transform for a certain group~\cite{Gottesman96,CRSS97} used in quantum
error-correcting.  Together with Beals' proposal of a quantum network
for the symmetric group~\cite{Beals97}, this emphasizes a challenging
question which has only been partly discussed in this paper.  Namely,
which applications are there for these new transforms?  Clearly, one can
define quantum versions of the classical applications, but are there any
other applications?  For example, a~crucial insight in Shor's
algorithm~\cite{Shor97} was the possibility of using the quantum version
of the discrete Fourier transform to find the index of an unknown
subgroup in a cyclic group.  No~efficient classical counterpart of this
idea is known.  Is~this phenomena present for non-Abelian groups, too?

\section*{Acknowledgments}
I~am very grateful to Joan Boyar and Gilles Brassard for many valuable
discussions and for their interest in this work.  I~am also grateful to
Andr\'e Berthiaume for interesting discussions on generalized Kronecker
products, and to Hans~J.\ Munkholm and Ren\'e Depont Christensen for
helpful discussions on representation theory.  This work was completed
at the Laboratoire d'informatique th\'eorique et quantique at
Universit\'e de Montr\'eal, and I~would like to thank the faculty and
the students, especially Alain Tapp, for their hospitality.

\end{document}